\documentclass[10pt]{iopart}

\pdfoutput=1

\usepackage{amssymb}
\usepackage{graphicx}
\usepackage{subfigure}

\usepackage{mathptmx}

\usepackage{hyperref}

\newcommand{\G}{\mathsf{G}}
\newcommand{\Pp}{\mathsf{p}}
\newcommand{\PP}{\mathsf{P}}
\renewcommand{\PM}{\mathsf{M}}
\newcommand{\Pm}{\mathsf{m}}
\newcommand{\mm}[2]{\Pm^{(#1)}_{#2}}

\newcommand{\rr}{\bi{r}}
\newcommand{\vv}{\bi{v}}
\newcommand{\ee}{\bi{e}}
\newcommand{\defeq}{\equiv}
\newcommand{\ii}{i}
\newcommand{\ds}{\scriptstyle}
\newcommand{\rin}{\rho_\mathrm{in}}
\newcommand{\rout}{\rho_\mathrm{out}}

\eqnobysec

\begin{document}
\title[Diffusive properties of persistent walks on cubic lattices]
{Diffusive properties of persistent walks on cubic lattices with
  application to periodic Lorentz gases}
\author{Thomas Gilbert\dag, Huu Chuong Nguyen\dag, David P. Sanders\ddag}
\address{\dag Center for Nonlinear Phenomena and Complex Systems, Universit\'e
  Libre  de Bruxelles, C.~P.~231, Campus Plaine, B-1050 Brussels, Belgium\\
  \ddag Departamento de F\'isica, Facultad de Ciencias, Universidad Nacional
  Aut\'onoma de M\'exico, 04510 M\'exico D.F., Mexico} 

\ead{thomas.gilbert@ulb.ac.be, hnguyen@ulb.ac.be, dps@fciencias.unam.mx}

\begin{abstract}
We calculate the diffusion coefficients of persistent
random walks on cubic and hypercubic lattices, where the direction of a
walker at a given step depends on the memory of one or two previous
steps. These results are then applied to study a billiard model,
namely a three-dimensional periodic Lorentz gas.
The geometry of the model is studied in order to find the regimes in 
which it exhibits normal diffusion. In this regime, we calculate numerically
the transition probabilities between cells to compare the 
persistent random-walk approximation with simulation results for the
diffusion coefficient. 
\end{abstract}

\submitto{J.~Phys.~A: Math.~Theor.}

\section{Introduction}

Problems dealing with the persistence of motion of tracer particles -- that
is, the tendency to continue or not in the same direction at a scattering
event -- are encountered in many areas of physics; see e.g.\
\cite{Haus:1987p513} and references therein. We are specifically interested
in the effect of persistence for the motion of random walkers on regular
lattices.

The diffusive properties of persistent random walks
on two-dimensional regular lattices were the subject of a previous paper by
two of the present authors \cite{Gilbert:2010p3733}. There, we presented a
theory making use of the symmetries of such lattices to derive the
transport coefficients of walks with a two-step memory. In the first part
of the present paper, we extend this theory to hyper-cubic lattices in
arbitrary dimensions, which is possible by describing the geometry of the 
lattices in a suitable way. 

Persistence effects naturally arise in the context of \emph{deterministic
diffusion} \cite{Geisel:1982p566, Fujisaka:1982p7971, Schell:1983p7988,
Grassberger:1983p563}, 
which is concerned with the interplay between dynamical properties at the
microscopic scale and transport properties at the macroscopic scale. A
variety of different techniques are now available, which rely on the
chaotic properties of model systems to describe their macroscopic
properties \cite{Gaspard:1998book},
\cite[chap.~25]{Cvitanovic:2004p284}. In particular, periodic Lorentz gases
and related models, such as multi-baker maps, are simple deterministic
dynamical systems  with strong chaotic properties which also exhibit
diffusive regimes. Although the transport
coefficients of these models can be expressed formally in terms of the
microscopic dynamical properties,
actually computing them is usually  difficult, with the exception
of some of the simplest toy models \cite{DorfmanBook1999, Klages:1995p649,
  Klages:1999p633}. One reason for this is 
that memory effects can remain important, in  spite of the chaotic
character of the underlying dynamics. 

The diffusive properties of these models therefore provide ideal
applications of the formalism presented in this paper.
An example of this was illustrated in reference \cite{Gilbert:2009p3207},
for a class of two-dimensional periodic billiard tables. 
Extending these results, in this paper we apply the formalism to model 
the diffusive properties of \emph{higher-dimensional} periodic Lorentz
gases.

The diffusive properties of the three-dimensional periodic Lorentz gas,
which consists of the free motion of independent tracer particles in a
cubic array of spherical obstacles, are interesting in their own right.
In two spatial dimensions, the existence of diffusive regimes in such
systems has been rigorously established \cite{Bunimovich:1980p2706,
  Bunimovich:1981p479}. It  relies on the \emph{finite-horizon} property,
which requires that the system admits no ballistic trajectories, i.e.\
those which never collide with any obstacle. In this case, it is possible
to change scales from microscopic to macroscopic, reducing the complicated
motion of tracer particles at the microscopic level to a diffusive equation
at the macroscopic level. When the horizon is infinite on the other hand,
there is rather a weakly superdiffusive process, with mean-squared
displacement growing like $t \log t$ \cite{Zacherl:1986p7768,
  Bleher:1992p315}, as recently shown rigorously in \cite{Szasz:2007p59}.  

The necessity of finite horizon to have normal diffusion in two dimensions
led to the idea that this was also necessary in three dimensions -- see,
for example, reference \cite{Chernov:1994p7678}.  
Recently, however, it was argued by one of the present authors 
\cite{Sanders:2008p453} that in
higher-dimensional billiards, normal diffusion, by which we mean an
asymptotically linear growth in time of the mean-squared displacement,  
may arise even in the \emph{absence} of finite
horizon. In fact, three different types of horizon can be identified in the
three-dimensional periodic Lorentz gas. The key observation is that 
it is only ``planar'' gaps -- those with infinite extension in two
dimensions --  which induce anomalous diffusion. If there are only
``cylindrical'' gaps, whose extension is limited to a single
dimension, then the available space in which particles can move
ballistically is limited. 
This leads to a decay of correlations which is fast enough to give normal
diffusion at the level of the mean-squared displacement, although higher
moments of the displacement distribution may be non-Gaussian
\cite{Sanders:2008p453}. 

The paper is organized as follows. \Sref{sec.PRW} describes the computation
of the transport coefficient of walks on hypercubic lattices with one and
two-step memories. In the second part of this paper, we apply this
formalism to the diffusive regimes of the three-dimensional periodic
Lorentz gas. In \sref{sec.LG3D}, we give a detailed description of the 
three-dimensional periodic Lorentz gas introduced in reference
\cite{Sanders:2008p453}, in particular delimiting the regimes with
qualitatively different behaviour in parameter space.  We then apply the
results on persistent random walks to the diffusive regimes of
this model in \sref{sec.num}. Conclusions are drawn in \sref{sec.con}.

\section{Persistent random walks on cubic lattices \label{sec.PRW}}

In this section, we describe a way to incorporate the specific geometry of
cubic and hyper-cubic lattices in the framework presented in
reference \cite{Gilbert:2010p3733} for calculating diffusion coefficients 
for persistent random walks on lattices.

We start by considering the motion of independent walkers on a regular
cubic lattice in three dimensions. Given their initial position $\rr_0$ at
time $t=0$, the walkers' trajectories are specified by the sequence
$\{\vv_{0},\ldots, \vv_{n}\}$ of their successive displacements. Here we  
consider dynamics in discrete time, so that the time sequences are simply
assumed to be incremented by identical time steps $\tau$ as the walkers
move from site to site. In the sequel we will loosely refer to the
displacement vectors as velocity vectors; they are in fact 
dimensionless unit vectors. 

The sequence of successive displacements is determined by the
underlying dynamics, whether deterministic or stochastic. At the coarse
level of description of the lattice dynamics, this is interpreted as a
\emph{persistent} type of random walk, where some memory effects are accounted
for: the probability that the $n$th step is taken in the direction $\vv_n$ depends on the
past history  $\vv_{n-1}, \vv_{n-2},\ldots$.

The quantity of interest here is the diffusion coefficient $D$ of
such persistent processes, which measures the linear growth in time of the
mean-squared displacement of walkers. This can be written in terms of
velocity autocorrelations using the Taylor--Green--Kubo expression:
\begin{equation}
  D = \frac{\ell^2}{2 d \tau} \left[1 + 2 \lim_{k\to\infty}
    \sum_{n = 1}^k \langle
    \vv_0 \cdot \vv_{n} \rangle \right]\,,
  \label{dcoef}
\end{equation}
where $d$ denotes the dimensionality of the lattice, here $d=3$, and $\ell$
is the lattice spacing. The (dimensionless) velocity autocorrelations
are computed as averages $\langle \cdot \rangle$ over the 
equilibrium distribution, denoted by $\mu$, of the underlying process, so
that the problem reduces to computing the quantities
\begin{equation}
  \langle \vv_0 \cdot \vv_{n} \rangle
  = \sum_{\vv_0, \ldots, \vv_{n}}  \vv_0 \cdot \vv_{n} \, 
  \mu(\{\vv_{0}, \ldots, \vv_{n}\}) \, .
  \label{v0vn}
\end{equation}

Following the approach of reference \cite{Gilbert:2010p3733}, we wish to
compute  the terms in this sum, and hence the corresponding diffusion
coefficient \eref{dcoef}, for three different types of random walks, namely
those with zero-step, single-step and two-step memories. These cases all
involve factorisations of the measure $\mu(\{\vv_{0}, \ldots, \vv_{n}\})$
into products of probability measures  which depend on a number of velocity
vectors, equal to the number of steps of memory of the walkers. These
measures will be denoted by $p$ throughout.

The schemes we outline below allow to write equation \eref{v0vn} as a sum
of powers of matrices, so that \eref{dcoef} boils down to a geometric
series, which can then be resummed to obtain an expression for the
diffusion coefficient that is readily computable given the probabilities
that characterise the allowed transitions in the process.

\subsection{Description of geometry of cubic lattices}
It is first necessary to find a succinct description of the geometry of the cubic lattices
that we wish to study.
The six directions of the three-dimensional cubic lattice and corresponding
displacement vectors are specified in terms of the unit vectors $\ee_i$
of a Cartesian coordinate system as $\pm \ee_i$, $i = 1,2,3$. 

The
crucial property required for the application of our method is
that all of these 
unit vectors can be obtained by repeated application of a single 
transformation $\G$, which generates the cyclic group 
\begin{equation}
  \mathcal{G} \defeq \{\G_i \defeq \G^i,\quad i=0,\ldots,5\}.
\end{equation}
One possible choice of $\G$ gives the following group elements:
\begin{equation}
\eqalign{
  \G_1 = - \G_4 = \G = 
    \left(
      \begin{array}{ccc}
        0 & 0 & -1 \\
        1 & 0 & 0 \\
        0 & 1 & 0
      \end{array}
    \right), 
     \\
    \G_2 = -\G_5 = \G^2 =
    \left(
      \begin{array}{ccc}
        0 & -1 & 0 \\
        0 & 0 & -1 \\
        1 & 0 & 0
      \end{array}
    \right), 
     \\
    \G_3 = - \G_0 = \G^3 =
    \left(
      \begin{array}{ccc}
        -1 & 0 & 0 \\
        0 & -1 & 0 \\
        0 & 0 & -1
      \end{array}
    \right).
   }
  \label{G3dcubic}
\end{equation}

\begin{figure}[htb]
  \centering
  \includegraphics[width=.5\textwidth,angle=0]{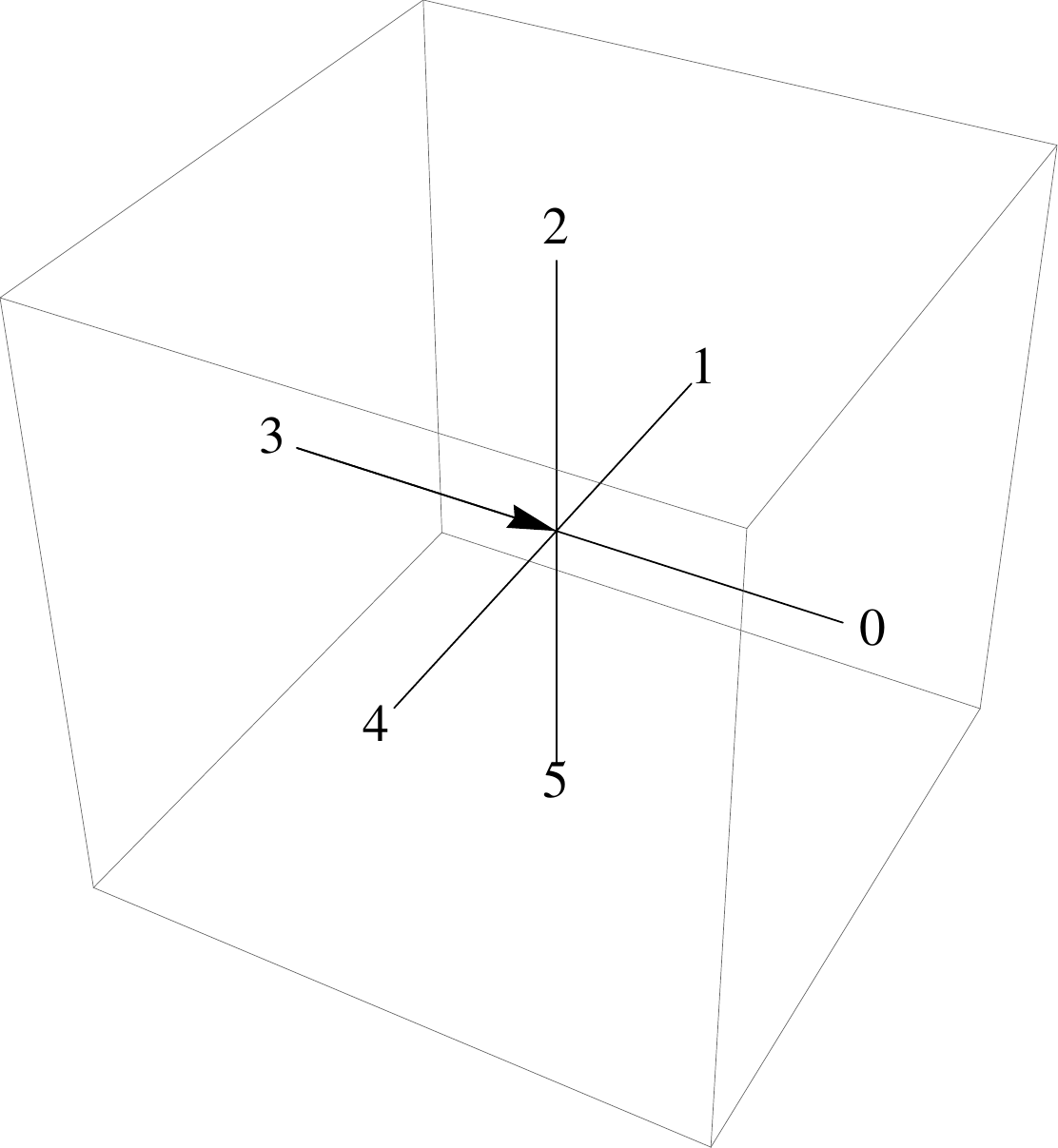}
  \caption{The possible directions of motion on a cubic lattice, labelled 
    from $0$ to $5$ \emph{relative} to the incoming direction 
    shown by the arrow. These directions are obtained by
    successive applications of the transformation $\G$ given in
equation~\eref{G3dcubic}.
    }
  \label{fig.num}
\end{figure}

\Fref{fig.num} displays the six possible directions of a
walker on this lattice, numbered according to repeated iterations by
$\G$. Thus a walker with incoming direction $\ee_1$, indicated by the
arrow, can be deflected to any of the six directions $\G^i \ee_1$,
$i=0,\ldots,5$, corresponding respectively to $\ee_1$, $\ee_2$,
$\ee_3$, $-\ee_1$, $-\ee_2$, and $-\ee_3$. 

A similar transformation $\G$ can easily be identified for a walk on
a $d$-dimensional hyper-cubic lattice:
\begin{equation}
  \G = 
  \left(
    \begin{array}{ccccc}
      0 & 0 & \cdots & 0 & -1\\
      1 & 0 & \cdots & 0 & 0\\
      0 & 1 & \cdots & 0 & 0\\
      \vdots&&\ddots&&\vdots\\
      0 & 0 & \cdots & 1 & 0
    \end{array}
  \right),
\end{equation}
which maps the unit vectors onto the $2d$-cycle $\ee_1 \mapsto
\ee_2 \mapsto \cdots \mapsto \ee_d \mapsto -\ee_1
\mapsto \cdots \mapsto -\ee_d$.

\subsection{\label{sec.nma}No-Memory Approximation (NMA)}
We now proceed to calculate the diffusion coefficient \eref{dcoef} for random walks with different memory lengths.
The simplest case is that of a Bernoulli process for the velocity
trials, so that the walkers have no memory of their history as they proceed to
their next position. The probability measure $\mu$ thus factorises: 
\begin{equation}
  \mu(\{\vv_{0},\ldots, \vv_{n}\}) = \prod_{i = 0}^n
  p(\vv_{i})\,.
\end{equation}
Given that the lattice is rotation invariant and that $p$ is uniform, the velocity
autocorrelation (\ref{v0vn}) vanishes:
\begin{equation}
  \langle \vv_{0} \cdot \vv_{n} \rangle = \delta_{n,0}\,.
\end{equation}
The diffusion coefficient of the random walk without memory is then given
by 
\begin{equation}
  D_\mathrm{NMA} = \frac{\ell^2}{2d\tau}\,.
  \label{DNMA}
\end{equation}

\subsection{\label{sec.1sma}One-Step Memory Approximation (1-SMA)}

We now assume that the velocity vectors obey a Markov process, for which
$\vv_{n}$ takes on different values according to the
velocity at the previous step $\vv_{n-1}$.  We may then write
\begin{equation}
  \mu(\{\vv_{0},\ldots, \vv_{n}\}) = \prod_{i = 1}^n 
  P(\vv_{i}|\vv_{i-1}) p(\vv_{0})\,.
  \label{mu1sma}
\end{equation}
Here, $P(\vv' | \vv)$ denotes the one-step conditional
probability that the walker moves with displacement $\vv'$,
given that it made a displacement $\vv$ at the previous step.

Considering for definiteness the three-dimensional lattice and using the
elements of the group $\mathcal{G}$, we express each velocity vector
$\vv_k$ in terms of the first one, $\vv_0$, as $\vv_k = \G^{i_k}  
\vv_0$, where each $i_k\in\{0,\ldots,5\}$. Substituting this into the
expression for the velocity autocorrelation $\langle \vv_{0} \cdot \vv_{n}
\rangle$, equation \eref{v0vn}, we obtain, using the factorisation
\eref{mu1sma}, 
\begin{equation}
  \fl \sum_{\vv_{0},\ldots, \vv_{n}}
  \vv_{0} \cdot \vv_{n}  \, \prod_{i = 1}^n 
  P(\vv_{i}|\vv_{i-1}) p(\vv_{0})
  =  \sum_{i_0,\ldots,i_n = 1}^6
  \vv_{0} \cdot \G^{i_n} \vv_0 \,
  \Pm_{i_n,i_{n-1}} \cdots \Pm_{i_1,i_0} \Pp_{i_0}\,.
  \label{v0vn0}
\end{equation}
In this expression,
\begin{equation}
  \Pm_{i_n,i_{n-1}} \defeq P(\G^{i_n} \vv_0|\G^{i_{n-1}} \vv_0) 
  \label{def-M}
\end{equation}
are the elements of the stochastic matrix $\PM$ of the Markov chain
associated to the persistent random walk, and $\Pp_i \defeq p(\ee_i)$
are the elements of its invariant (equilibrium) distribution, denoted
$\PP$, evaluated with a velocity in the $i$th lattice direction. The
invariance of $\PP$ is expressed as $\sum_{j}\Pm_{i,j}\Pp_j = \Pp_i$. The
same notations were used in \cite{Gilbert:2010p3733} and will be used
throughout this article.

The terms involving $\PM$ in (\ref{v0vn0}) constitute the matrix product of
$n$ copies of $\PM$. Furthermore, 
since the invariant distribution is uniform over the lattice
directions,  we can choose an arbitrary direction for $\vv_0$, and hence
write 
\begin{eqnarray}
  \langle \vv_0 \cdot  \vv_n \rangle
  &= \vv_0 \cdot \vv_0 \, \mm{n}{1,1} 
  + \vv_0 \cdot \G \vv_0 \, \mm{n}{2,1}
  + \cdots
  + \vv_0 \cdot \G^5 \vv_0 \, \mm{n}{6,1}\,,
  \nonumber\\
  &= \mm{n}{1,1} - \mm{n}{4,1} 
  \label{v0vn1sma}
\end{eqnarray}
where $\mm{n}{i,j}$ denote the elements of $\PM^n$.

The actual value of the diffusion coefficient depends on the probabilities
$P(\G^j\vv|\vv)$, which are parameters of the model, subject to the
constraints $\sum_j P(\G^j\vv|\vv) = 1$. To simplify
the notation, we assume rotational invariance of the process, i.e.\
independence with respect to the value of $\vv$, and we denote the
conditional probabilities of these walks by 
$P_j \defeq  P(\G^j\vv|\vv)$, where $j = 0,\ldots, 5$.

The transition matrix $\PM$ given by (\ref{def-M})  is thus the cyclic matrix
\begin{equation}
  \PM = 
  \left(
    \begin{array}{c@{\enspace}c@{\enspace}c@{\enspace}c@{\enspace}c@{\enspace}c}
      P_0 & P_1 & P_2 & P_3 & P_4 & P_5\\
      P_5 & P_0 & P_1 & P_2 & P_3 & P_4\\
      P_4 & P_5 & P_0 & P_1 & P_2 & P_3\\
      P_3 & P_4 & P_5 & P_0 & P_1 & P_2\\
      P_2 & P_3 & P_4 & P_5 & P_0 & P_1\\
      P_1 & P_2 & P_3 & P_4 & P_5 & P_0
    \end{array}
  \right)\,.
  \label{P1SMA}
\end{equation}
The matrix $\PM^n$ shares the same property of cyclicity, so that it also
has only six distinct entries. It is thus possible to 
proceed along the lines described in \cite{Gilbert:2010p3733} and obtain
the recurrence relation 
\begin{eqnarray}
  \fl
  \left(
    \begin{array}{c}
      \mm{n}{1,1} - \mm{n}{4,1}\\
      \mm{n}{2,1} - \mm{n}{5,1}\\
      \mm{n}{3,1} - \mm{n}{6,1}
    \end{array}
  \right)
  &= 
  \left(
    \begin{array}{c@{\enspace}c@{\enspace}c}
      P_0 - P_3 & P_1 - P_4 & P_2 - P_5\\ 
      P_5 - P_2 & P_0 - P_3 & P_1 - P_4\\ 
      P_4 - P_1 & P_5 - P_2 & P_0 - P_3
    \end{array}
  \right)
  \left(
    \begin{array}{c}
      \mm{n-1}{1,1} - \mm{n-1}{4,1}\\
      \mm{n-1}{2,1} - \mm{n-1}{5,1}\\
      \mm{n-1}{3,1} - \mm{n-1}{6,1}
    \end{array}
  \right),\nonumber\\
  &= 
  \left(
    \begin{array}{c@{\enspace}c@{\enspace}c}
      P_0 - P_3 & P_1 - P_4 & P_2 - P_5\\ 
      P_5 - P_2 & P_0 - P_3 & P_1 - P_4\\ 
      P_4 - P_1 & P_5 - P_2 & P_0 - P_3
    \end{array}
  \right)^{n-1}
  \left(
    \begin{array}{c}
      P_0 - P_3\\
      P_1 - P_4\\
      P_2 - P_5
    \end{array}
  \right).
  \label{recmn1sma}
\end{eqnarray}
[Note that the left-hand side of this equation was chosen to reduce the
size of the matrix involved and to calculate the element required in
\eref{v0vn1sma}.] As a consequence, we can write for the velocity
autocorrelation 
(\ref{v0vn1sma})
\begin{equation}
  \fl 
  \eqalign
  \langle \vv_0\cdot\vv_n \rangle 
  = \big(1 \enspace 0 \enspace 0 \big)
  \left(
    \begin{array}{c@{\enspace}c@{\enspace}c}
      P_0 - P_3 & P_1 - P_4 & P_2 - P_5\\ 
      P_5 - P_2 & P_0 - P_3 & P_1 - P_4\\ 
      P_4 - P_1 & P_5 - P_2 & P_0 - P_3
    \end{array}
  \right)^{n-1}
  \left(
    \begin{array}{c}
      P_0 - P_3 \\
      P_1 - P_4 \\
      P_2 - P_5
    \end{array}
  \right),
  \label{v0vn1SMAsq} 
\end{equation}
and thus obtain the expression of the diffusion coefficient \eref{dcoef} as
\begin{equation}
  \fl
  \frac{D_\mathrm{1SMA}}{D_\mathrm{NMA}} = \left[1 +
    2 \Big(1\enspace 0 \enspace 0 \Big)
    \left(
      \begin{array}{c@{\enspace}c@{\enspace}c}
        \ds 1 + P_3 - P_0 & \ds P_4 - P_1 & \ds P_5 - P_2\\ 
        \ds P_2 - P_5 & \ds 1 + P_3 - P_0 & \ds P_4 - P_1\\ 
        \ds P_1 - P_4 & \ds P_2 - P_5 & \ds 1 + \ds P_3 - P_0
      \end{array}
    \right)^{-1}
    \left(
      \begin{array}{c}
        P_0 - P_3 \\
        P_1 - P_4 \\
        P_2 - P_5
      \end{array}
    \right)
  \right],
  \label{D1SMA}
\end{equation}
by using the result that $\sum_{n=0}^{\infty} \mathsf{A}^n = (\mathsf{I} -
\mathsf{A})^{-1}$, where $\mathsf{I}$ is the identity
matrix, for a square matrix $\mathsf{A}$ whose eigenvalues are
all strictly less than $1$ in modulus, 

This result easily generalises to a hyper-cubic lattice in any dimension
$d$. Note also that for a symmetric process, in which $P_1 = P_4$ and $P_2
= P_5$, we recover the diffusion coefficient
\begin{equation}
  D_\mathrm{1SMA} = D_\mathrm{NMA}\frac{1 + P_0 - P_3} {1 - P_0 + P_3},
    \label{D1SMASq}
\end{equation}
in agreement with the result stated in \cite{Gilbert:2010p3733}.

\subsection{\label{sec.2sma}Two-Step Memory Approximation (2-SMA)}

Let us now suppose that the velocity vectors obey a random process for which
the probability of $\vv_{n}$ takes on different values according to the
velocities at the two previous steps, $\vv_{n-1}$ and $\vv_{n-2}$, so that 
we may write
\begin{equation}
  \mu(\{\vv_{0},\ldots, \vv_{n}\}) = \prod_{i = 2}^n 
  P(\vv_{i}|\vv_{i-1}, \vv_{i-2}) p(\vv_{0},
  \vv_{1})\,.
\end{equation}
The velocity autocorrelation (\ref{v0vn}) function is then
\begin{equation}
  \langle   \vv_{0} \cdot \vv_{n} \rangle 
  = \sum_{\{\vv_{n},\ldots,\vv_{0}\}}
  \vv_{0} \cdot \vv_{n} \,
 \prod_{i = 2}^n 
  P(\vv_{i}|\vv_{i-1}, \vv_{i-2}) p(\vv_{0},
  \vv_{1})\,.
  \label{2spa0}
\end{equation}

Since the probability transitions $P(\vv_{i}|\vv_{i-1}, \vv_{i-2})$ have
symmetries similar to those used in reference \cite{Gilbert:2010p3733}, the
computation of equation \eref{2spa0} reduces to an expression very similar
to that found there for walks on one- and two-dimensional lattices. The details
of the derivation are a bit more involved than the one-step memory
persistent walks, so we will limit ourselves to stating the results.

Letting $z = 2d$ denote the coordination number of the lattice, and
writing\footnote{This expression differs from that given in
  \cite{Gilbert:2010p3733} due to a typographical error in that paper --
  they are really the same.} 
$P_{j,k} \defeq  P(\G^{z-k}\G^{z-j}\vv|\G^{z-j}\vv, \vv)$, which is  the
conditional probability of making a displacement $\vv$ given that the two
preceding displacements were successively $\G^j\G^k\vv$ and $\G^k\vv$, we
define the $z\times z$ matrix
\begin{equation}
  K(\phi) \equiv
    \left(
      \begin{array}{c@{\enspace}c@{\enspace}c@{\enspace}c}
        \ds P_{00} & \ds P_{10} & \ds \cdots & \ds P_{z-1,0}  \\
        \ds \phi P_{01} & \ds \phi P_{11} & \ds \cdots & \ds \phi
        P_{z-1,1} \\   
        \ds \vdots & \ds \vdots & \ds \ddots & \ds \vdots \\
        \ds \phi^{z-1} P_{0,z-1} & \ds \phi^{z-1} P_{1,z-1} & \ds \cdots
        & \ds  \phi^{z-1} P_{z-1,z-1}
      \end{array}
    \right).
    \label{K}
\end{equation}
The argument $\phi$ in this expression is a complex number such that
$\phi^z = 1$. In the case of  two-dimensional lattices, only two of
these roots are relevant, corresponding to the complex exponential of the
smallest angle between two lattice vectors, $\phi = \exp(\pm
2\ii\pi/z)$. For hyper-cubic lattices in arbitrary dimensions, however, we
must consider a priori all the $z$ possible roots of unity, $\phi_j \equiv
\exp(2\ii\pi j/z)$, $j = 0,\ldots, z-1$. 

A direct calculation of \eref{2spa0} shows that the velocity autocorrelation
takes the form  
\begin{equation}
  \fl
  \langle   \vv_{0} \cdot \vv_{n} \rangle
  = \Big( 1 \cdots 1\Big)\left[
    \sum_{j = 0}^{z-1} a_j K(\phi_j)^{n-1}
    \mathrm{diag}(1, \phi_j, \ldots, \phi_j^{z-1})
  \right]
  \left(
    \begin{array}{c}
      \Pp_1 \\
      \vdots \\
      \Pp_z
    \end{array}
  \right),
  \label{v0vn2SMA}
\end{equation}
where $\mathrm{diag}(1, \phi_j, \ldots, \phi_j^{z-1})$ denotes the
matrix with elements listed on the main diagonal and $0$ elsewhere.
For the three-dimensional cubic lattice, the coefficients $a_j$ are found
to be 
\begin{equation}
  \eqalign{
    a_0 = a_2 = a_4 = 0,\\
    a_1 = a_3 = a_5 = 2,
  }
\end{equation}
which compares to $a_1 = a_3 = 2$ and $a_0 = a_2 = 0$ in the case of the
two-dimensional square lattice \cite{Gilbert:2010p3733}. In the case of a
$d$-dimensional hyper-cubic lattice, this generalises to
\begin{equation}
  \eqalign{
    a_{2j} = 0,\quad j=0,\ldots, d-1,\\
    a_{2j+1} = 2,\quad j=0,\ldots, d-1,
  }
\end{equation}

The diffusion coefficient of a two-step memory persistent random walk on a
$d$-dimensional hyper-cubic lattice is thus
\begin{eqnarray}
  \fl
  \frac{D_\mathrm{2SMA}}{D_\mathrm{NMA}}
  = 1 + 4 \Big( 1 \cdots 1\Big)
    \left\{
      \sum_{j = 1}^{d} [I_z - K(\phi_{2j-1})]^{-1}
      \mathrm{diag}(1, \phi_{2j-1}, \ldots, \phi_{2j-1}^{z-1})
    \right\}
  \left(
    \begin{array}{c}
      \Pp_1 \\
      \vdots \\
      \Pp_z
    \end{array}
  \right),
  \nonumber\\
  \label{D2SMA}
\end{eqnarray} 
where $I_z$ denotes the $z\times z$ identity matrix.

\section{Three-dimensional periodic Lorentz gas \label{sec.LG3D}}

Equations \eref{DNMA}, \eref{D1SMA} and \eref{D2SMA} can be put to the test
to probe the diffusive regimes of periodic Lorentz gases. 
The diffusive
motion of the tracers results from the chaotic nature of the microscopic
dynamics and the fast decay of correlations, which are in turn due to the
convex nature of the obstacles. Taking into consideration the different
diffusive regimes of these models, which, as we argued earlier, depend on
the nature of their horizon, we  investigate how the microscopic dynamical
properties of the system determine the diffusion coefficient. 

Machta and Zwanzig \cite{Machta:1983p182} addressed this issue in a 
particular limiting case, showing that, in the limit where the obstacles
are so close together that a tracer will remain localised on ekach lattice site
for a very long time (compared to the mean time separating two collision
events), the process of diffusion on the Lorentz gas is well approximated
by the dimensional prediction \eref{DNMA}, where the lattice 
spacing $\ell$ is the distance separating two neighbouring obstacles and
$\tau$ is the trapping time, which can be computed in terms of the geometry
of the billiard as a simple consequence of ergodicity. That is to say, when
the geometry of the billiard is such that two neighbouring disks nearly
touch, the Lorentz gas is well approximated by a Bernoulli process,
modeling the random hopping of tracers from cell to cell, with time- and
length-scales specified according to the geometry of the billiard.

Different approximation schemes have been proposed to go beyond this 
zeroth-order approximation and account for corrections to
it \cite{Klages:2000p146, Klages:2002p264}; see, in particular, reference
\cite{Klages:2007p202} for an overview. A consistent approach 
to understanding the effect of these corrections in two-dimensional
diffusive billiards was described in \cite{Gilbert:2009p3207}. The idea is
to approximate the hopping process of tracer particles by persistent random
walks with finite memory, and thus estimate the diffusion coefficient of
the billiard by the two-dimensional lattice equivalents of the one- or
two-step formulas \eref{D1SMA} and \eref{D2SMA}.

We discuss below the transposition of these results to the diffusive
regimes of the three-dimensional periodic Lorentz gas.

\subsection{Geometry of simple three-dimensional periodic 
Lorentz gas model} 

We begin with a detailed description of the geometry and the different horizon
regimes of the system studied in reference \cite{Sanders:2008p453};
additional details are given in reference \cite{Sanders:2008p315}. 

\begin{figure}[htb]
  \centering
  \includegraphics[scale=0.2]{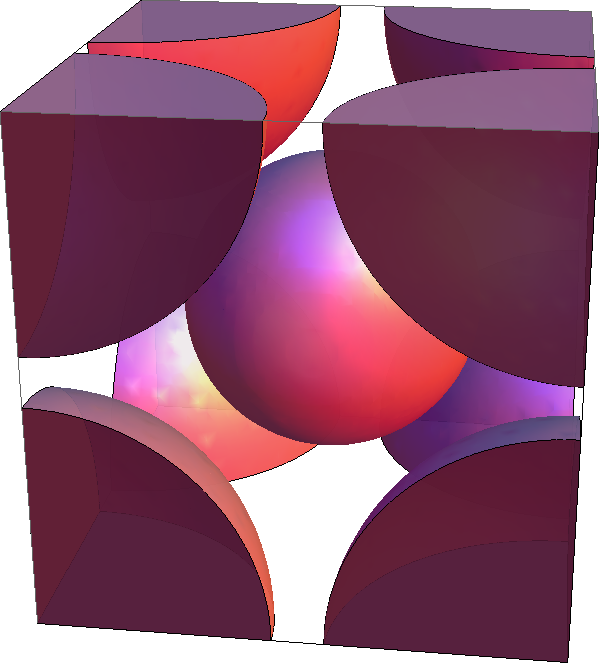}
  \caption{Geometry of the obstacles in a single cell of the 3D periodic
    Lorentz gas model for $\rout = 0.45 \ell$ and $\rin = 0.30 \ell$, in
    the cylindrical-horizon regime.} 
  \label{fig.lg}
\end{figure}

The model consists of a three-dimensional (3D) periodic Lorentz gas
constructed out of cubic unit cells of side length $\ell$, having eight
``outer'' spheres of radius $\rout\ell$ at its corners and a single
``inner'' sphere of radius $\rin\ell$ at its centre -- see
\fref{fig.lg}. The infinitely-extended periodic structure formed in this
way is symmetric under interchange of $\rin$ and $\rout$; without loss of
generality, we take $\rout \ge \rin$. 

This model seems to be the simplest one which allows a finite horizon,
although this is possible only when the  spheres are permitted to
overlap. It is
known that finite-horizon periodic Lorentz gases with non-overlapping
spheres  in fact exist in any dimension \cite{Henk:2000p31}, but we are not
aware of any explicit constructions of such models, even in the case of
three dimensions.

\paragraph{Lattice of outer spheres}
The spheres of radius $\rout \ell$ form a simple cubic lattice. This
lattice has the following properties: 
\begin{itemize}
 \item 
   When $\rout < 1/2$, the spheres are disjoint. In this case, there
   are \emph{free planes} \cite{Henk:2000p31} in the structure, that is,
   infinite planes which  do not intersect any of the spheres, in
   particular there are free planes centered on the faces of the unit cell.
   In this case, we say that there is a \emph{planar horizon} (PH). When
   $\rout$ is small, there are additional planes at different diagonal
   angles, analogously to the two-dimensional infinite-horizon Lorentz gas
   \cite{Zacherl:1986p7768, Bleher:1992p315, Szasz:2007p59}.   
\item 
  When $\rout > 1/2$, the spheres overlap, thereby automatically blocking
  all planes. The overlaps (intersections) of the spheres partially cover
  the faces of the cubes, leaving a space in between which acts as an exit
  towards the adjacent cell. 
\item 
  When  $\rout \ge 1/\sqrt{2}$, the
  overlaps completely cover the faces of the unit cell, so that it is no
  longer possible to exit the cell. 
\item 
  When $\rout \ge \sqrt{3}/2$, all of space is covered, and it is no longer
  possible to define a billiard dynamics. 
\end{itemize}

\paragraph{Conditions for normal diffusion: cylindrical horizon}
As shown in  reference \cite{Sanders:2008p453}, the necessary and
sufficient condition to have normal diffusion  is that all free planes are
blocked; if there are free planes, then the diffusion is weakly
anomalous.  The conditions to block all planes are as follows.
\begin{itemize}
\item 
  All free planes are automatically blocked for $\rout \ge 1/2$, when
  the $\rout$-spheres overlap. 
\item 
  If the $\rout$-spheres do not overlap, then it is necessary to introduce the
  $\rin$-sphere to block planes which are parallel to the faces of the unit
  cell. For this blocking to occur, we need $\rin \ge 1/2 - \rout$.   
\item
  Furthermore, we must also block diagonal planes at $45$ degree angles,
  which requires that $\rout \ge 1/(2\sqrt{2})$ or $\rin \ge 1/(2\sqrt{2})$.
\end{itemize}  
If all of these conditions are satisfied, then
we no longer have free planes, but may have free cylinders (``cylindrical
gaps'') in the structure; we then say that there is a \emph{cylindrical
  horizon} (CH).

\begin{figure}[tbh]
  \centering
  \subfigure[Face of unit cell]{
    \includegraphics[scale=0.5]{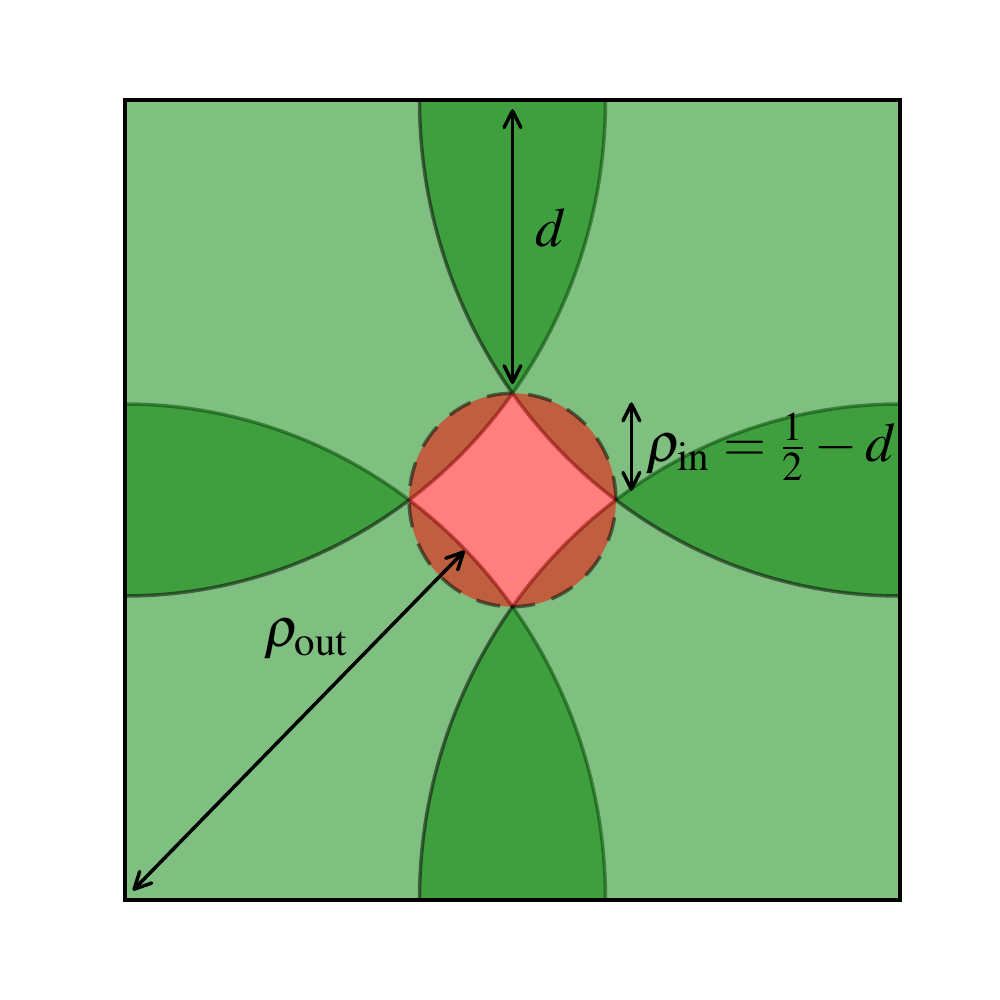}
    \label{fig.overlap-face}
  }
  \quad
  \subfigure[Mid-plane of unit cell]{
    \includegraphics[scale=0.5]{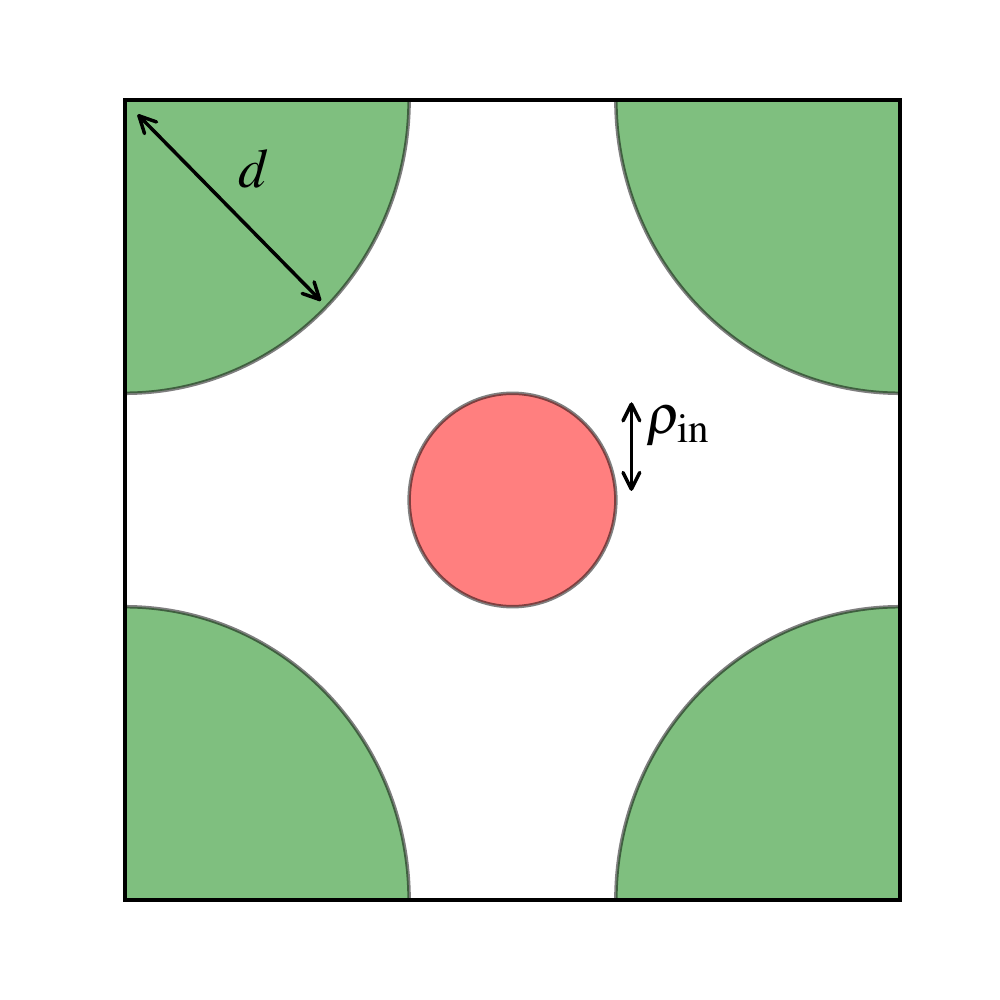}
    \label{fig.overlap-mid-plane}
  }
  \caption{Geometry of the 3D periodic Lorentz gas. (a) Cross-section of
    the unit cell in one of its faces. The overlapping outer spheres of
    radius $\rout > 1/2$, give rise to four overlapping discs (shown in
    green); the  maximum width of their overlap is denoted $d$. The central
    disc (red) shows the minimum radius     $\rin = 1/2 - d$ of the central
    sphere such that its projection covers the gap between the
    $\rout$-discs on the face. (b) Geometry of the mid-plane of a unit cell
    for parameters giving a finite horizon. The outer discs are
    cross-sections of the overlaps of the outer $\rout$-spheres, and have
    radius $d$ equal to the overlap parameter in (a). The inner disc is the
    cross-section of the inner $\rin$-sphere.
  }
  \label{fig.geometry}
\end{figure}

\begin{figure}[htb]
  \centering
  \includegraphics[scale=0.4]{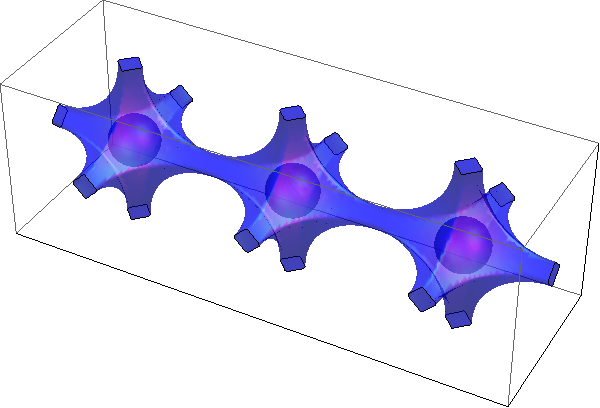}
  \caption{Finite horizon can be achieved in a three-dimensional lattice,
    provided the spheres are allowed to overlap. Here the available space
    for diffusing particles is shown for parameter values $\rout =
    0.65 \ell$ and $\rin = 0.15 \ell$ (in the FH1 region) in an unfolded channel.
  } 
  \label{fig.lgchannel}
\end{figure}
\paragraph{Conditions for finite horizon}
Stronger statistical properties -- e.g.\ faster decay of correlations --
may be expected when there is a \emph{finite} horizon
\cite{Chernov:1994p7678, Sanders:2008p453}, i.e.\ where the length of free
paths between collisions with obstacles is bounded above. 
To obtain this, not only all planar gaps, but also all cylindrical gaps
must be blocked, i.e.\ all holes viewed from any direction must be
blocked. To do so, the following conditions must be fulfilled: 
\begin{itemize}
\item
  The $\rout$-spheres must overlap, $\rout \ge 1/2$. 
  Furthermore, the projection of the $\rin$-sphere 
  on each face of the unit cell must cover the available exit space, as
  illustrated in \fref{fig.overlap-face}. Letting $d$ be the maximum width of
  overlap of the resulting discs of radius $\rout$ on a face of the unit
  cell, we have $d^2 = \rout^2 - 1/4$, and we need $\rin \ge 1/2 - d$ to
  block the space. 
\item
  We must block cylindrical corridors which cross the structure at a
  $45$ degree angle  at the level of the \emph{mid-plane} of a unit
  cell, which corresponds to the planar cross-section with most available
  space in the unit cell. The mid-plane has the geometry shown in
  \fref{fig.overlap-mid-plane}, with four outer discs of radius $d$, and a
  central disc of radius $\rin$; these discs are the intersection of the
  $\rout$-overlaps and of the $\rin$-sphere, respectively, with the mid-plane. Free diagonal
  trajectories in this plane at an angle of $45$ degrees give rise to small
  cylindrical corridors.  These will be blocked if there is no free line in
  the mid-plane. This blocking occurs  provided either $d
  \ge 1/(2 \sqrt{2})$, i.e.\ 
  $\rout \ge \sqrt{3} / (2 \sqrt{2}$, 
or
  if $\rin \ge 1/(2 \sqrt{2})$, thus giving rise to two distinct finite
  horizon regimes (FH1 and FH2), which are in fact disjoint.
\end{itemize}
\Fref{fig.lgchannel} depicts the space available for tracer particles  
in a channel of three consecutive cells for a particular
finite-horizon case.  

\paragraph{Localisation of trajectories}
Having fixed $\rout$, it is also necessary to calculate the value of
$\rin$ above which the trajectories become localised (L) between
neighbouring spheres, and are thus no longer able to diffuse. For 
$\rout < 1/\sqrt{2}$, when there are still exits available on the
faces of the cubic unit cell, this happens exactly when the discs in the
mid-plane touch, i.e.\ when $\rin + d = 1/\sqrt{2}$, so that the
condition for localised trajectories becomes \cite{Sanders:2008p453} $\rin
\ge 1/\sqrt{2} - \sqrt{\rout^{2} - 1/4}$. 

\paragraph{Condition to fill space}
Finally, we calculate when the spheres fill all space (denoted U, for
undefined):
\begin{itemize}
\item
  When $\rout < 1/\sqrt{2}$, this occurs when the $\rin$-spheres are
  large enough that their intersection with each face of the cube, which
  is a disc of radius $\sqrt{\rin^2 - 1/4}$, covers the exit on a
  face left open by the $\rout$-spheres. This gives the condition $\rin^2
  \ge \rout^2 + 1/4 - \sqrt{\rout^2 - 1/4}$. 
\item
  When $\rout > 1/\sqrt{2}$, the condition is that $\rin$ be large
  enough to cover the space left by the $\rout$-spheres inside the unit
  cell. The condition can again be found by looking at the mid-plane, where
  there is most available space: the disc of radius $\rin$ must cover the
  space left by the discs of radius $d$ (which are cross-sections of the
  overlaps of the $\rout$-spheres). This occurs when $\rin \ge 1/2 -
  \sqrt{\rout^2 -1/2}$.  
\end{itemize}

\begin{figure}[thb]
  \centering
  \includegraphics[scale=0.7]{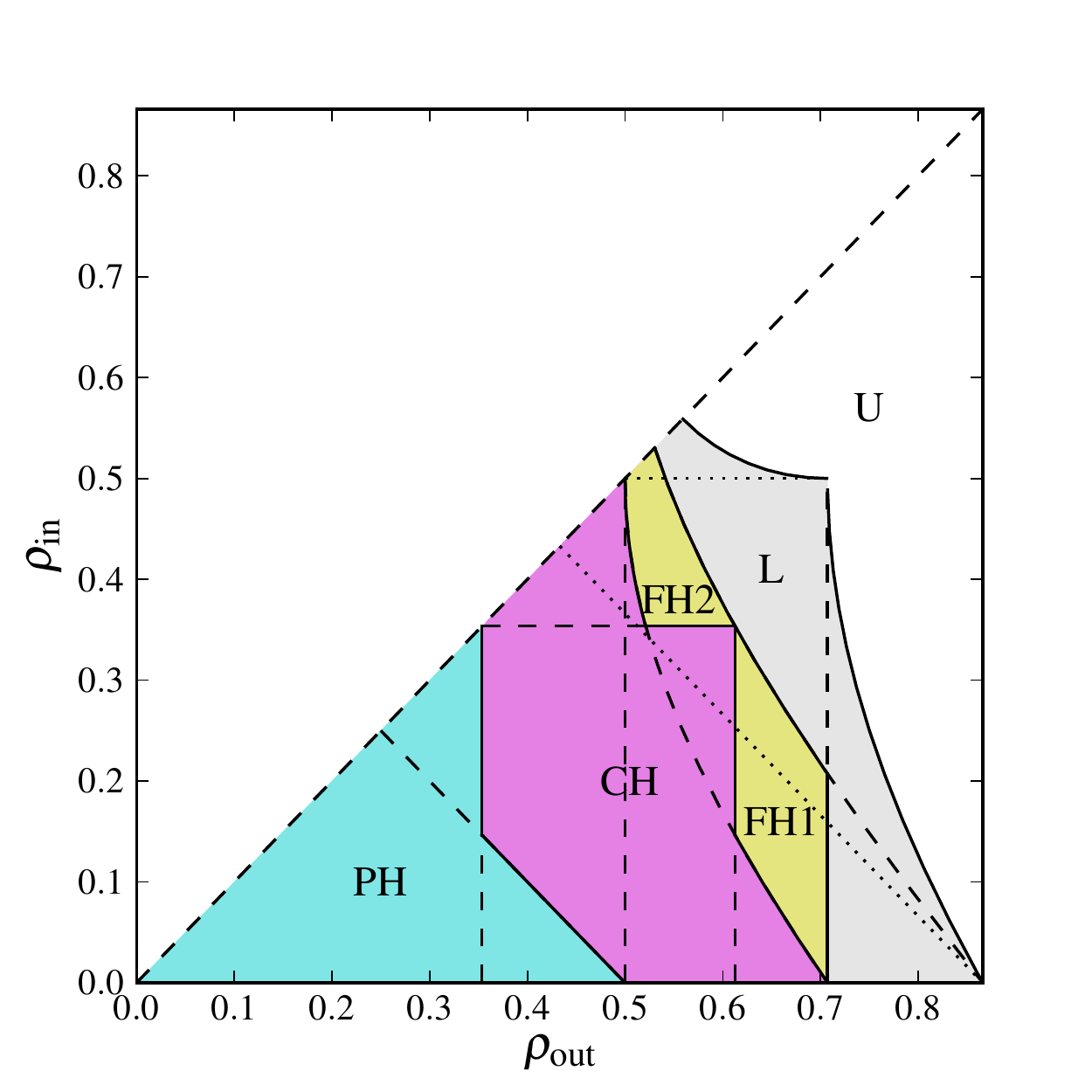}
  \caption{Parameter space of the three-dimensional periodic Lorentz gas as
    a function of the geometrical parameters $\rout$ and $\rin$. Solid
    lines divide regimes of qualitatively different behaviour, which are
    also shaded with different colours and labelled as follows: PH: planar
    horizon; CH: cylindrical horizon; FH1 and FH2: finite horizon; L:
    localised, non-diffusive motion; U: undefined (all space filled). Note
    that the FH regime is divided into two disjoint regions. The dashed
    lines mark the different conditions referred to in the text.  The
    diagonal dotted line separates regions where the $\rin$-spheres do
    (above) and do not (below) overlap the $\rout$-spheres. The diagram is
    reflection-symmetric in the line $\rin = \rout$, but for clarity only
    the lower half is shown.
  }
  \label{fig.phase-diag}
\end{figure}

\paragraph{Parameter space}
The complete parameter space of this model is shown in
\fref{fig.phase-diag}, exhibiting the regions in parameter space
corresponding to the regimes of qualitatively different behaviour discussed
above%
\footnote{A similar diagram of parameter
  space for a two-dimensional version of the model was given in
  reference \cite{Garrido:1997p807}. However, the symmetry
  between $\rout$ and $\rin$ was overlooked there; see also 
  reference \cite{Sanders:2008p453}.}.  
Note that if $\rin > \sqrt{3}/2 - \rout$, then the $\rout$- and
$\rin$-spheres overlap, and if $\rin > 0.5$ then neighbouring
$\rin$-spheres also overlap. These conditions are marked by the 
dotted lines in the figure.

\section{Persistence in the diffusive regimes of the three-dimensional
  Lorentz gas \label{sec.num}} 

In this section, we study the dependence of the diffusion coefficient on 
the geometrical parameters of the 3D periodic Lorentz gas model in the
finite- (FH1) and cylindrical-horizon (CH) regimes, comparing the numerical
results with the finite-memory approximations \eref{DNMA}, \eref{D1SMA} and
\eref{D2SMA}.

\subsection{Approximation by the NMA process}

The computation of the dimensional formula \eref{DNMA} relies on that of
the residence time $\tau$. An exact formula is available for
this quantity \cite{Chernov:1997p1}:
\begin{equation}
  \tau = \frac{|Q|}{|\partial Q|}\frac{|S^2|}{|B^2|},
  \label{tau3d}
\end{equation}
where $|Q|$ denotes the volume of the billiard domain outside the obstacles,
$|\partial Q|$ the surface
area of the available gaps separating neighbouring cells, $|S^2| = 4\pi$ 
the surface area of the unit sphere in three dimensions, and $|B^2| = \pi$
the volume (area) of the unit disk in two dimensions, and we assume unit
velocity. The explicit formulas giving the values of $|Q|$ and $|\partial
Q|$ are rather lengthy and will not be given here; see 
reference \cite{Nguyen:2010p7979}.  

\begin{figure}[thb]
  \centering
  \includegraphics[width=.5\textwidth,angle=0]{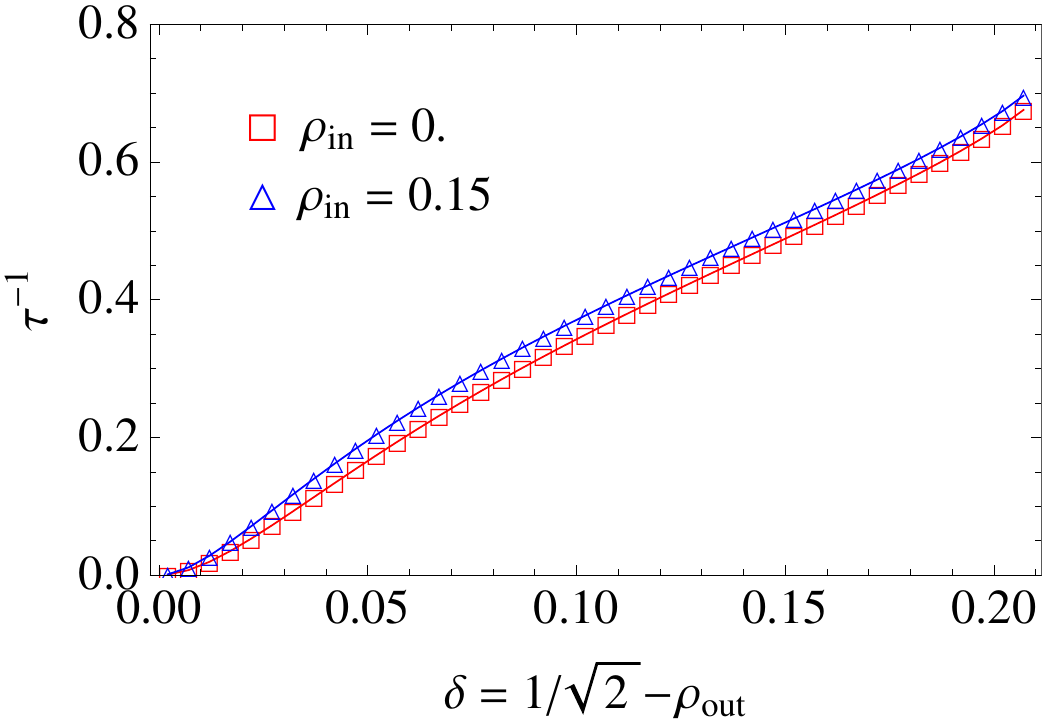}
  \caption{Residence time $\tau$, equation~\eref{tau3d}, compared to direct
    numerical simulations. The results are shown for two values of 
    the inner radius, $\rin = 0$ and $\rin = 0.15$, as functions of $\delta
    \defeq 1/\sqrt{2} - \rout$, which is the characteristic size of the gaps 
    separating neighbouring cells. The curves are very similar
    since the volume of the inner
    sphere remains small. In this and the following results we take
    $\ell = 1$.
  }
  \label{fig.tau}
\end{figure}

The validity of equation \eref{tau3d} can be tested by comparison with
numerical computation of the residence time, as shown in \fref{fig.tau}.
Here, and in the remainder of the paper, we restrict attention to values of
$\rout$ close to the limiting value $1/\sqrt{2}$ and $\rin$ close
to $0$, so that the  geometry is that of a single, cubic unit cell.

\subsection{Approximation by the 1SMA and 2SMA processes}

Single- and two-step memory processes can be derived as approximations, at the
lattice level, to the dynamics of the Lorentz gas. This is  done by
computing numerically the statistics of tracer particles as they jump from
cell to cell, so as to estimate the single- and two-step memory probability
transitions.

\begin{figure}[htbp]
  \centering
  \subfigure[]{
    \includegraphics[width=.3\textwidth,angle=0]{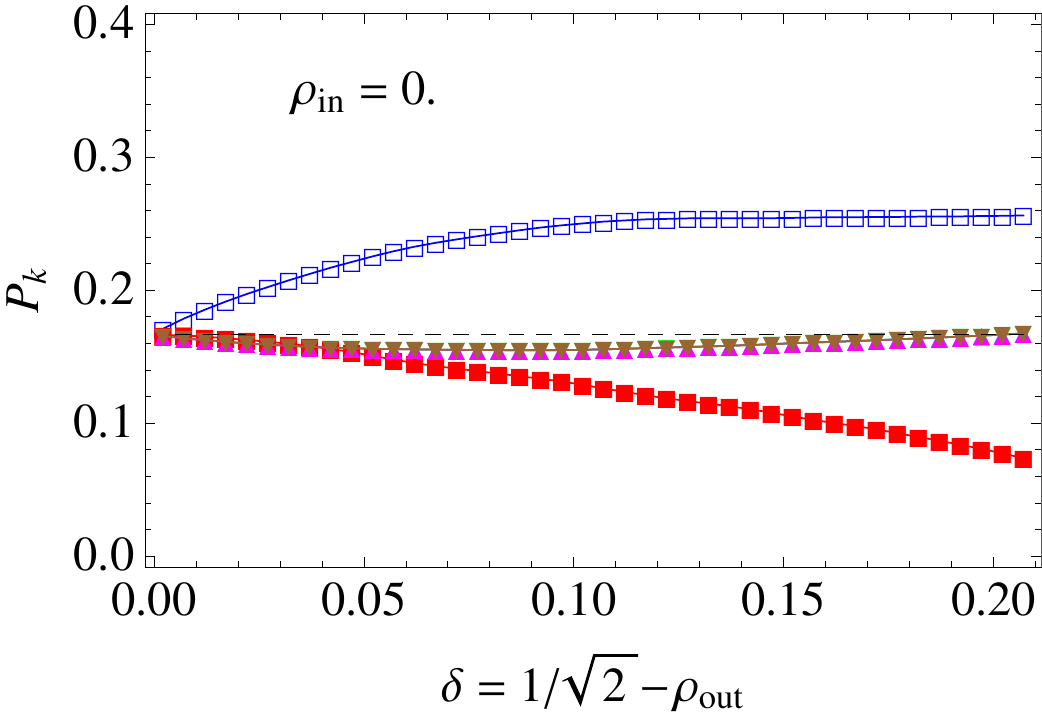}
  }
  \hfill
  \subfigure[]{
    \includegraphics[width=.3\textwidth,angle=0]{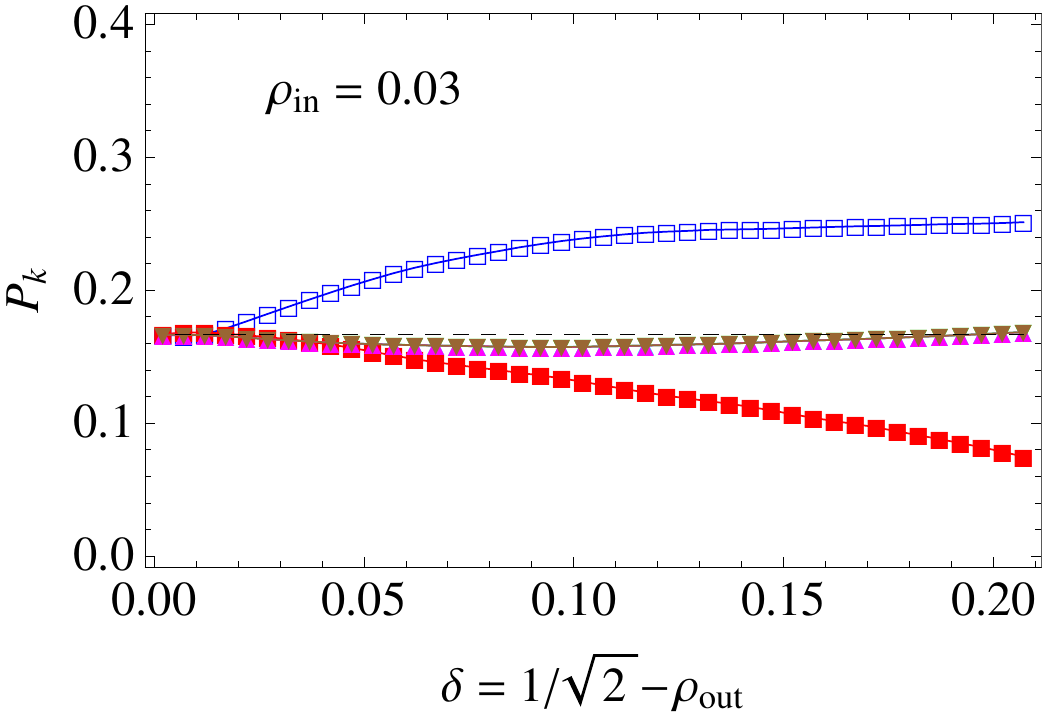}
  }
  \hfill
  \subfigure[]{
    \includegraphics[width=.3\textwidth,angle=0]{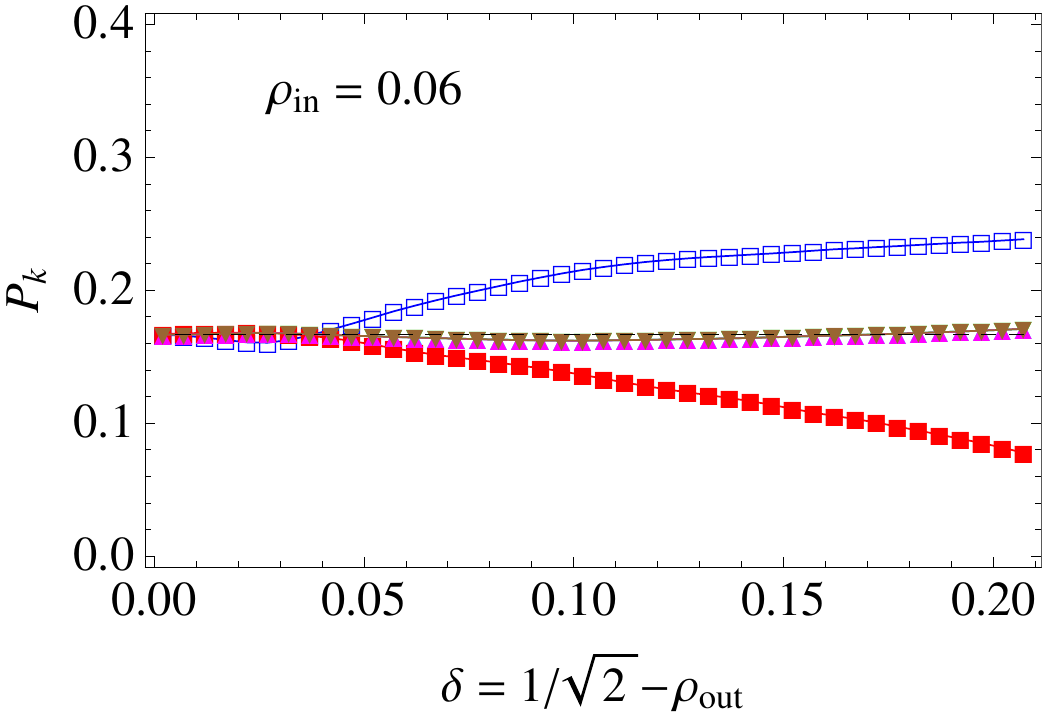}
  }
  \\
  \hfill
  \subfigure[]{
    \includegraphics[width=.3\textwidth,angle=0]{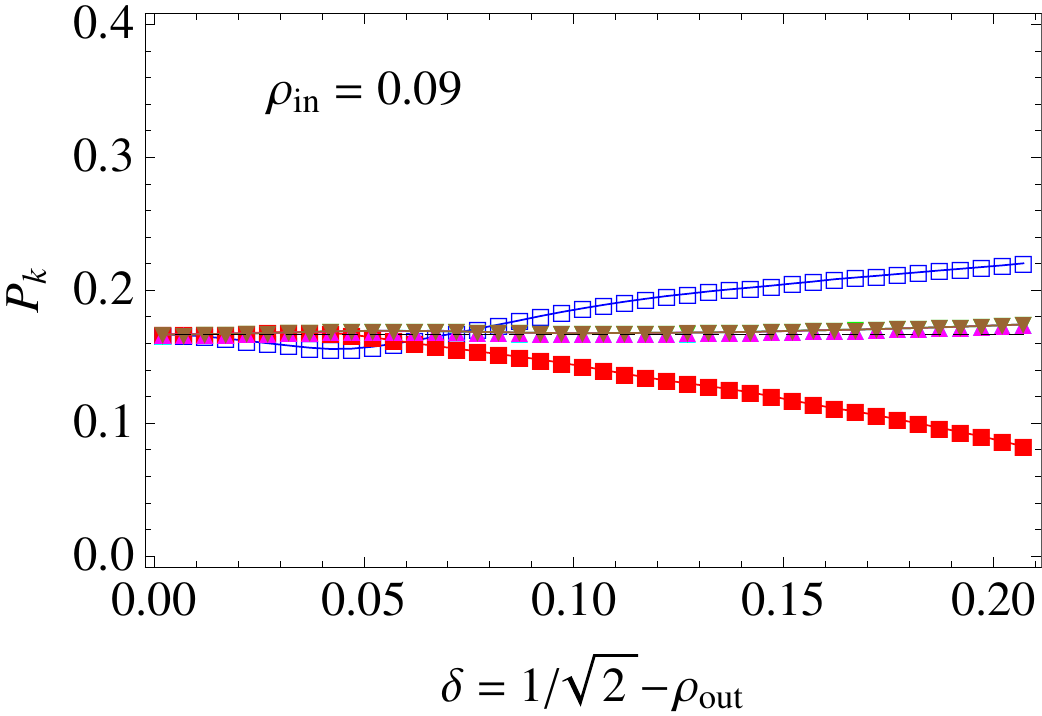}
  }
  \hfill
  \subfigure[]{
    \includegraphics[width=.3\textwidth,angle=0]{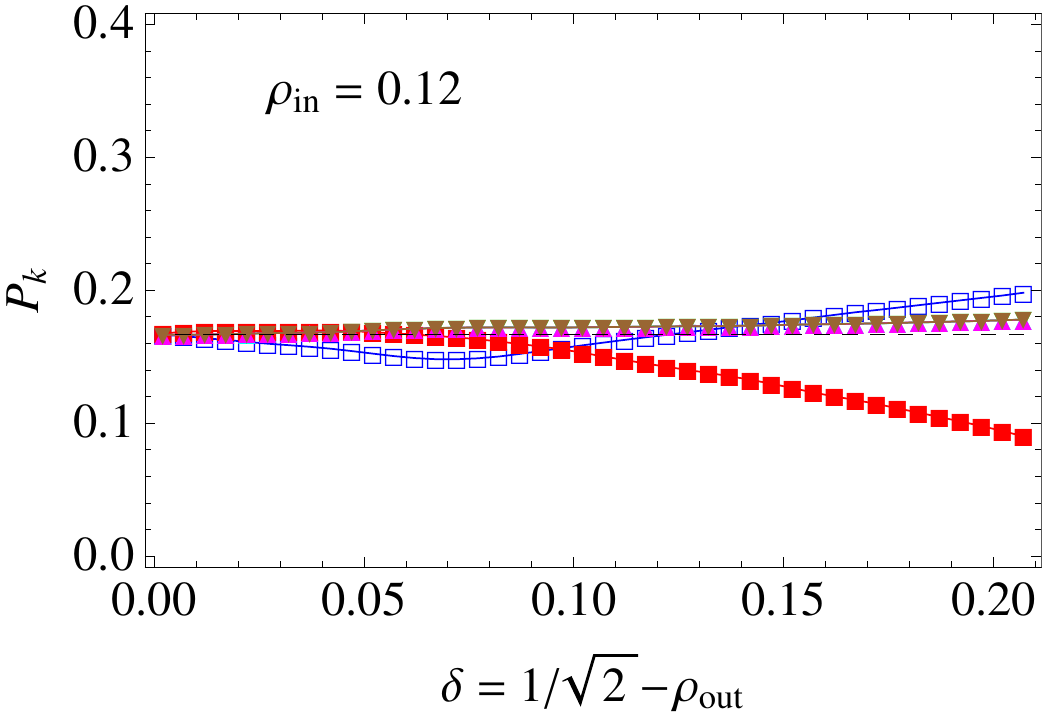}
  }
  \hfill
  \subfigure[]{
    \includegraphics[width=.3\textwidth,angle=0]{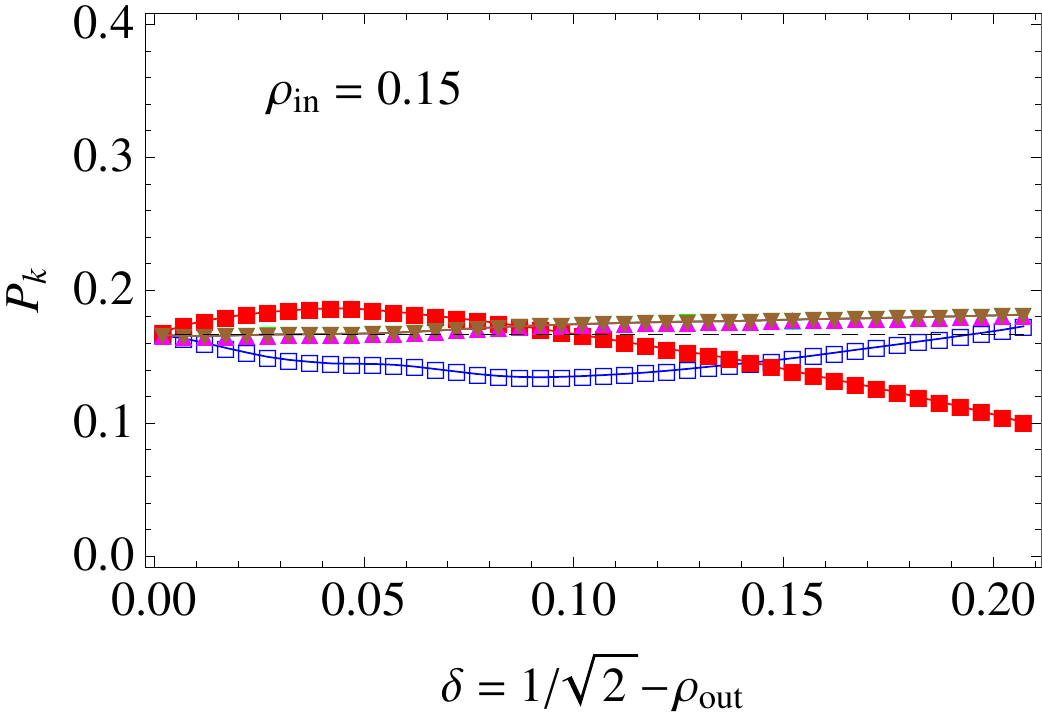}
  }
  \caption{Numerical computations of the probabilities $P_0,\ldots,P_5$
    of the single step memory process, appearing in \eref{P1SMA}. The
    six panels shown correspond to as many different values of $\rin$,
    where the probabilities are shown as functions of $\delta$. The dashed
    line at $P_k = 1/6$ indicates the   
    value for a memoryless (NMA) walk. Here and in figures \ref{fig.p2_b00}
    and \ref{fig.p2_b15}, the conventions are as follows: Empty squares
    (blue), $P_0$; empty upward triangles (cyan), $P_1$; empty downward
    triangles (green), $P_2$; 
    filled squares (red) $P_3$; filled upward triangles (magenta), $P_4$;
    filled downward triangles (brown), $P_5$. In all cases we verify the
    symmetry $P_1 = P_2 = P_4 = P_5$, which also remain close to $1/6$.
  }
  \label{fig.p1}
\end{figure}

The results are shown in \fref{fig.p1} for the single-step memory process,
where the six transition probabilities $P_i$, $i=0,\ldots,5$, are displayed
as functions of the outer radius $\rout$ for different values of the inner
radius $\rin$.  

\begin{figure}[htbp]
  \centering
  \subfigure[]{
    \includegraphics[width=.3\textwidth,angle=0]{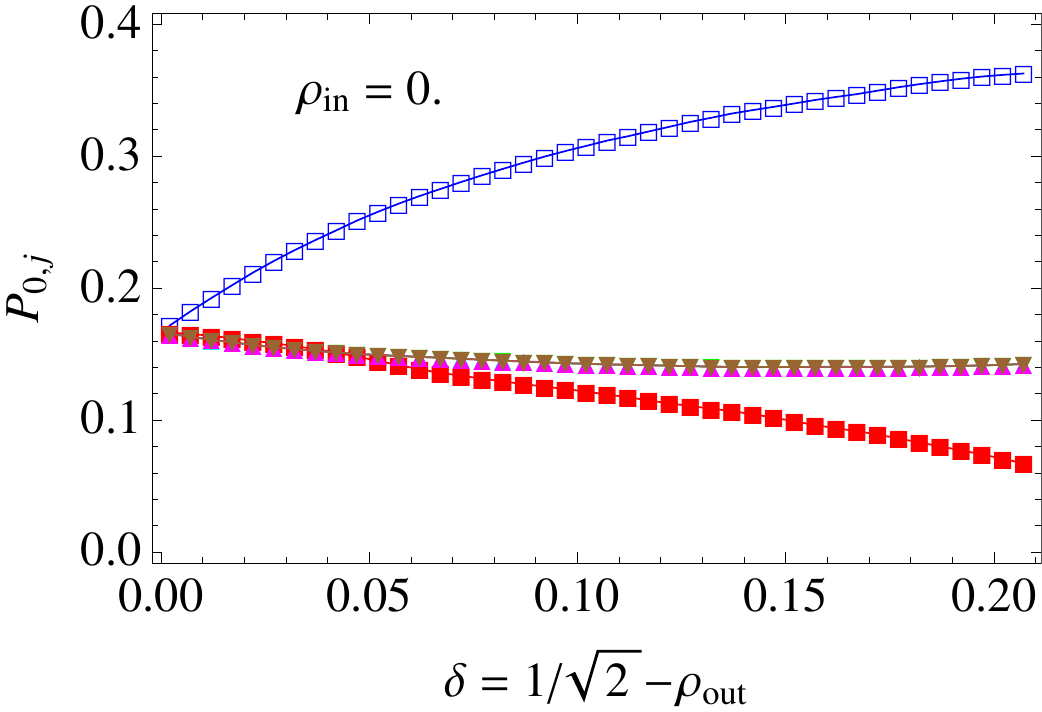}
    \label{fig.p20_b00}
  }
  \hfill
  \subfigure[]{
    \includegraphics[width=.3\textwidth,angle=0]{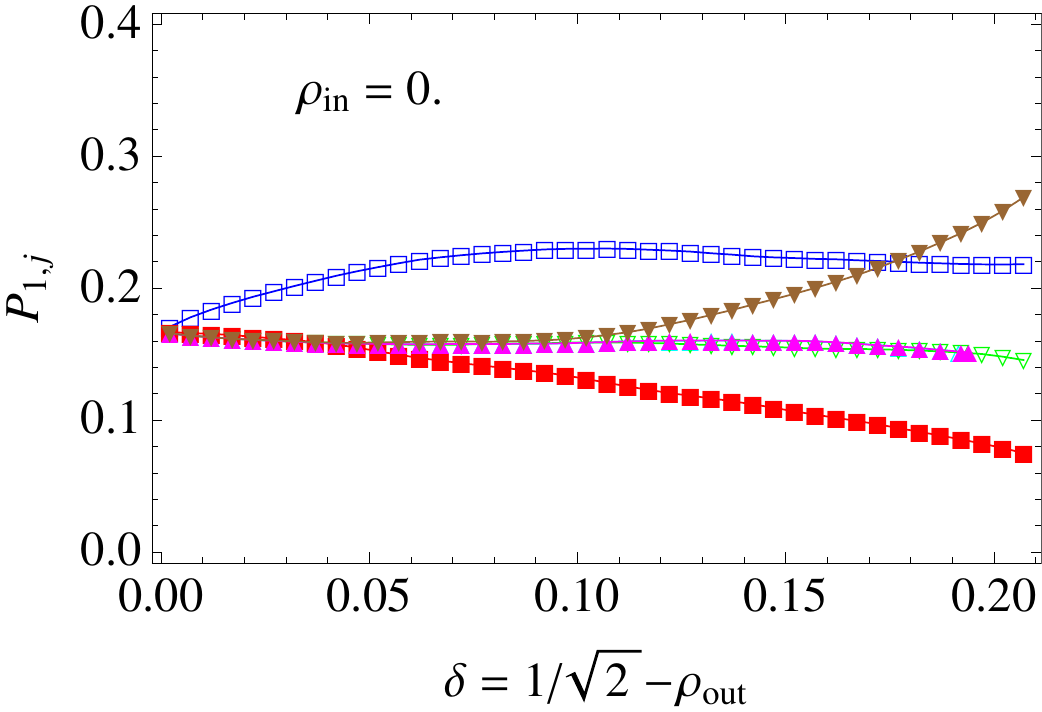}
    \label{fig.p21_b00}
  }
  \hfill
  \subfigure[]{
    \includegraphics[width=.3\textwidth,angle=0]{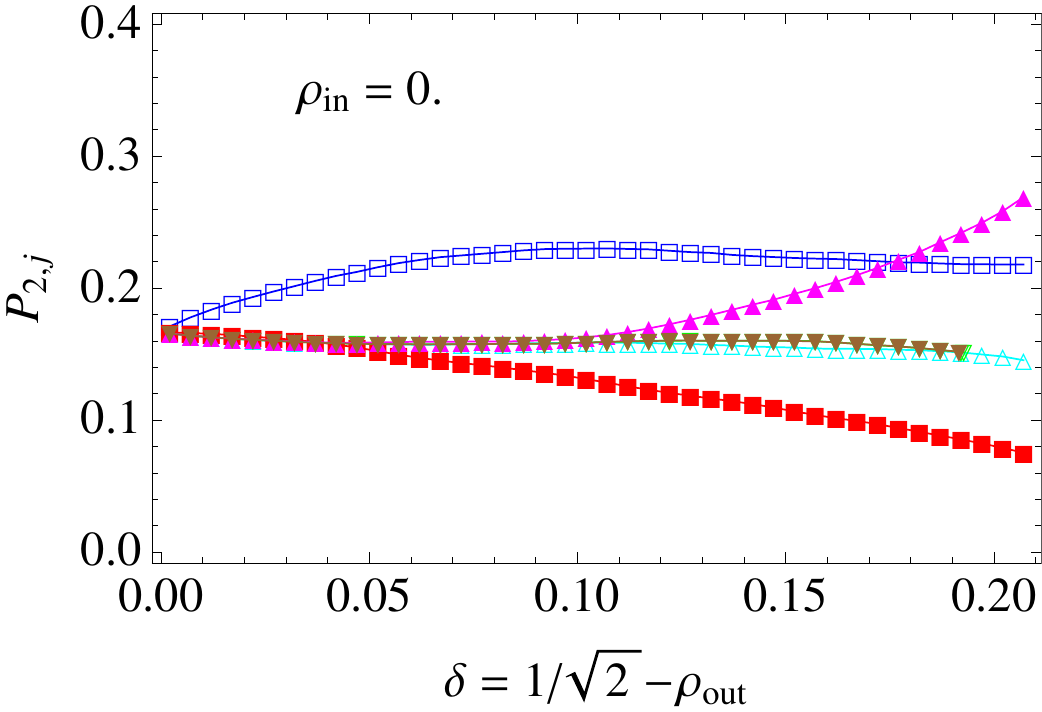}
    \label{fig.p22_b00}
  }
  \\
  \subfigure[]{
    \includegraphics[width=.3\textwidth,angle=0]{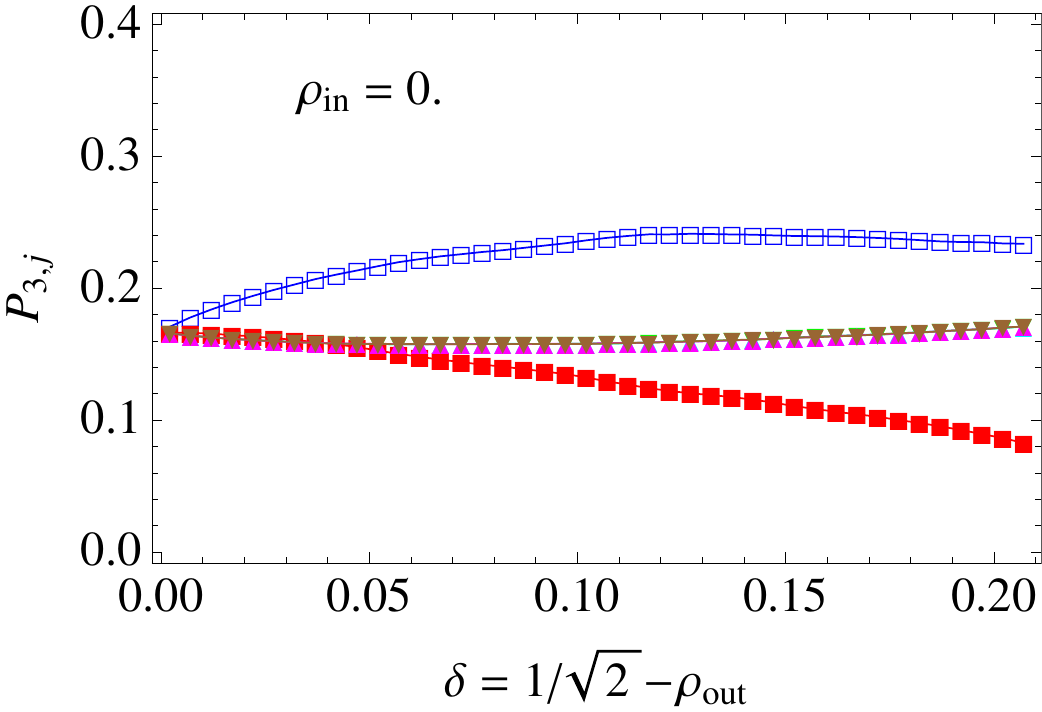}
    \label{fig.p23_b00}
  }
  \hfill
  \subfigure[]{
    \includegraphics[width=.3\textwidth,angle=0]{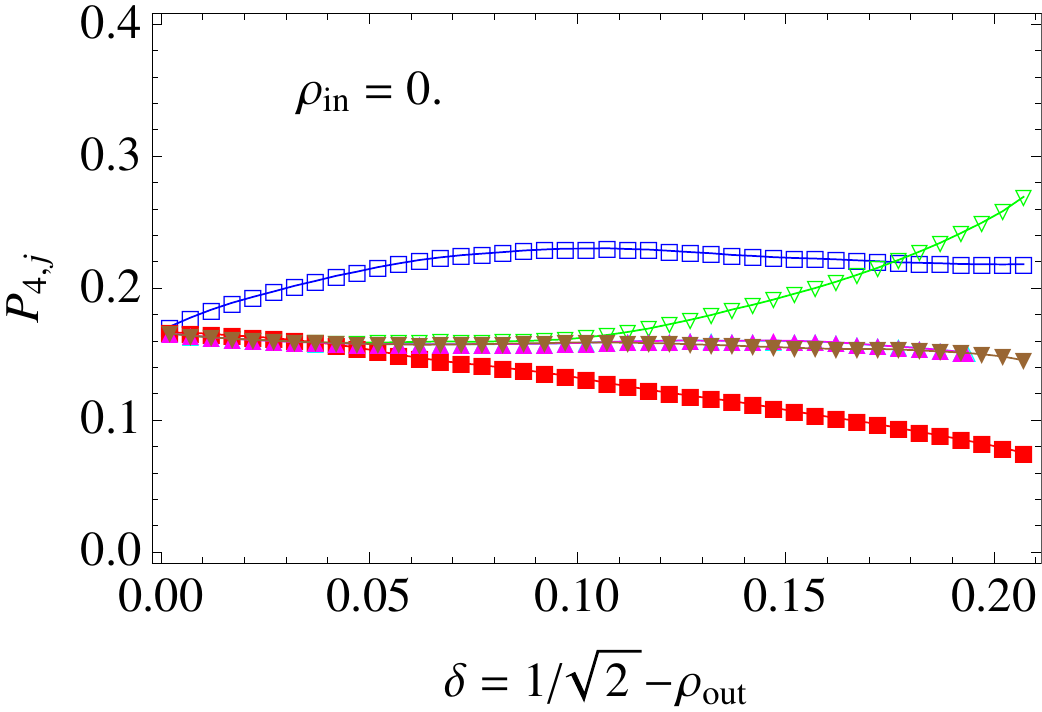}
    \label{fig.p24_b00}
  }
  \hfill
  \subfigure[]{
    \includegraphics[width=.3\textwidth,angle=0]{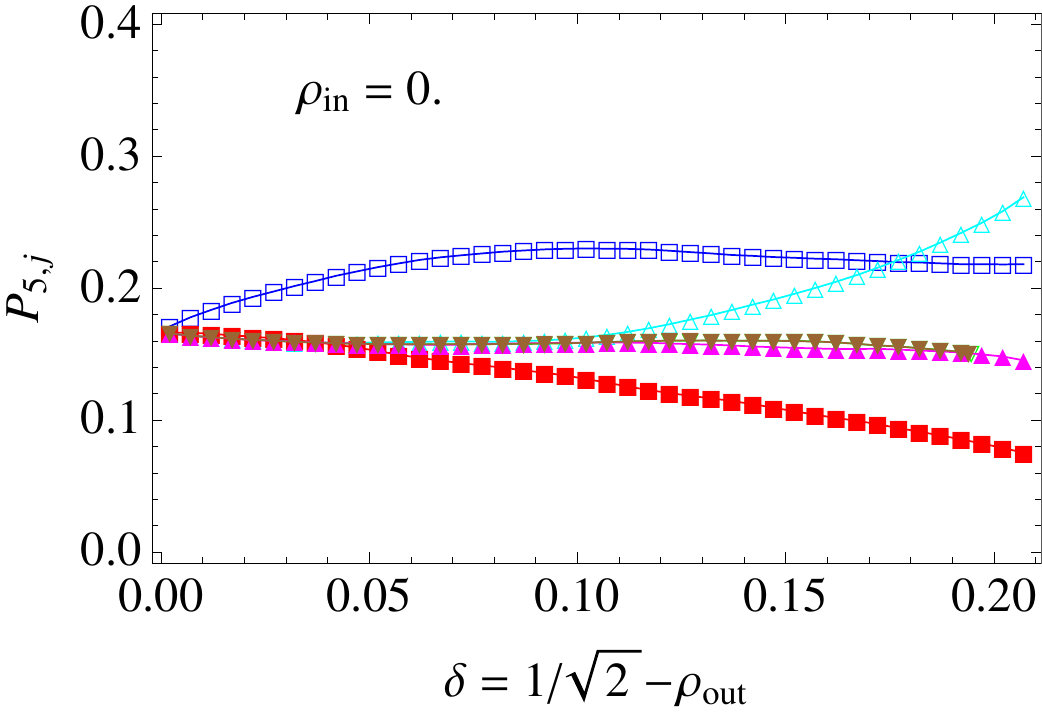}
    \label{fig.p25_b00}
  }
  \caption{Numerical computations of the 36 probabilities $P_{i,j}$
    which appear in \eref{K}, corresponding to a cell with no sphere at its
    center, i.e.\ the inner radius $\rin = 0$. The symmetries of the
    process are reflected by the similarities between figures
    \ref{fig.p21_b00}, \ref{fig.p22_b00}, \ref{fig.p24_b00} and
    \ref{fig.p25_b00}. The colour coding is similar to \fref{fig.p1}.
  }
  \label{fig.p2_b00}
\end{figure}

\begin{figure}[htbp]
  \centering
  \subfigure[]{
    \includegraphics[width=.3\textwidth,angle=0]{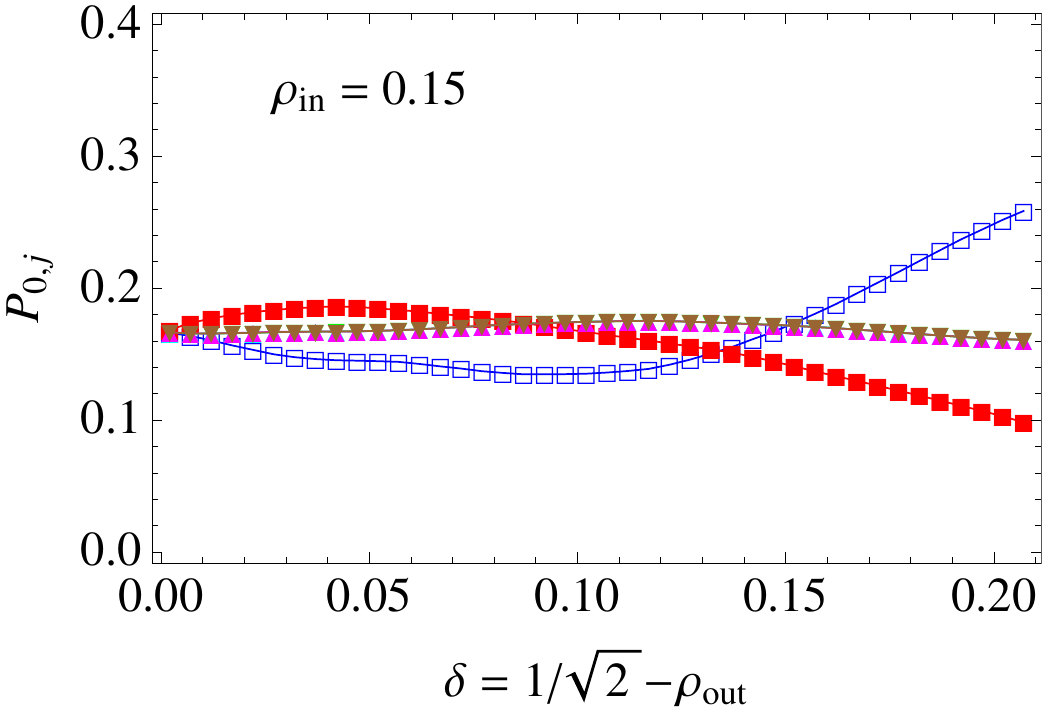}
  \label{fig.p20_b15}
}
  \hfill
  \subfigure[]{
    \includegraphics[width=.3\textwidth,angle=0]{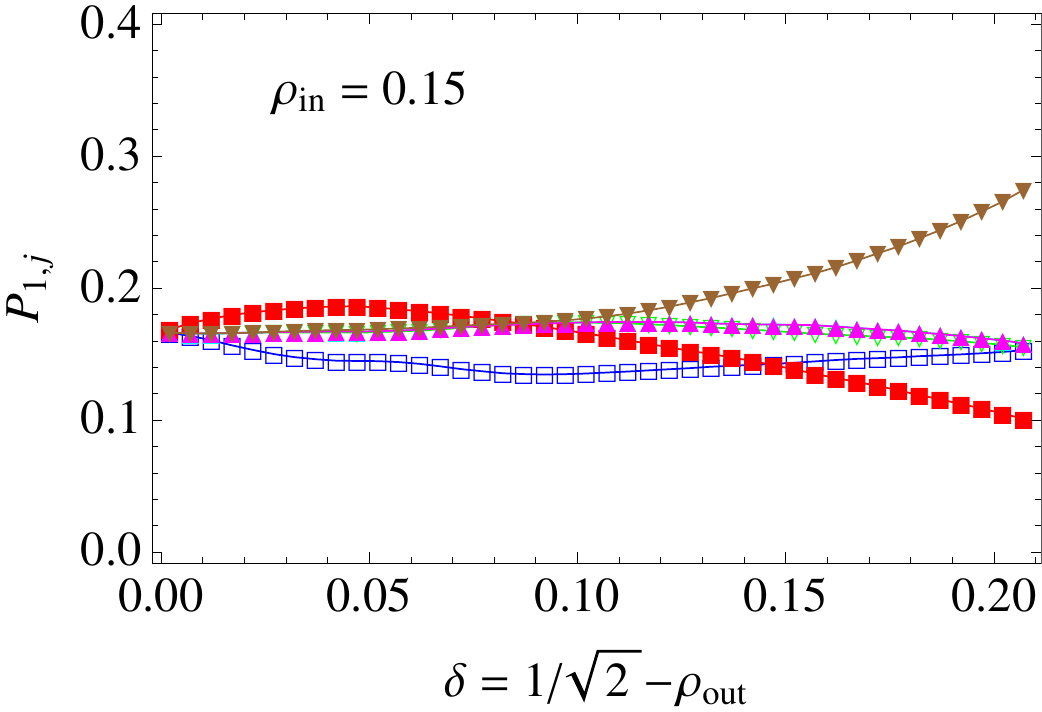}
  \label{fig.p21_b15}
  }
  \hfill
  \subfigure[]{
    \includegraphics[width=.3\textwidth,angle=0]{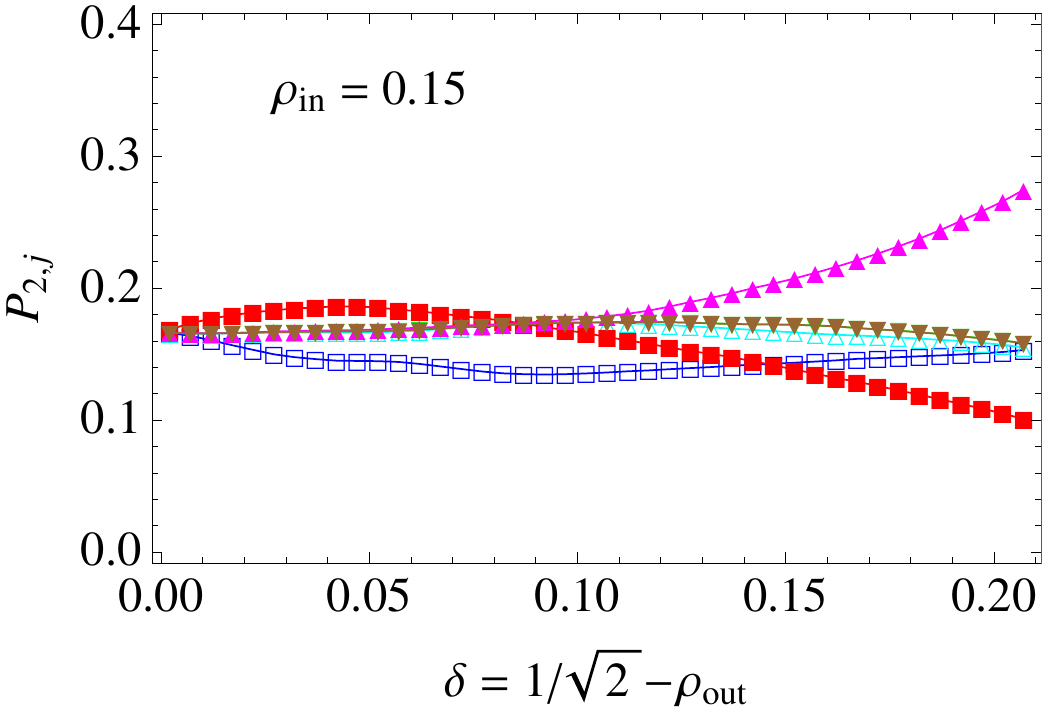}
  \label{fig.p22_b15}
  }
  \\
  \subfigure[]{
    \includegraphics[width=.3\textwidth,angle=0]{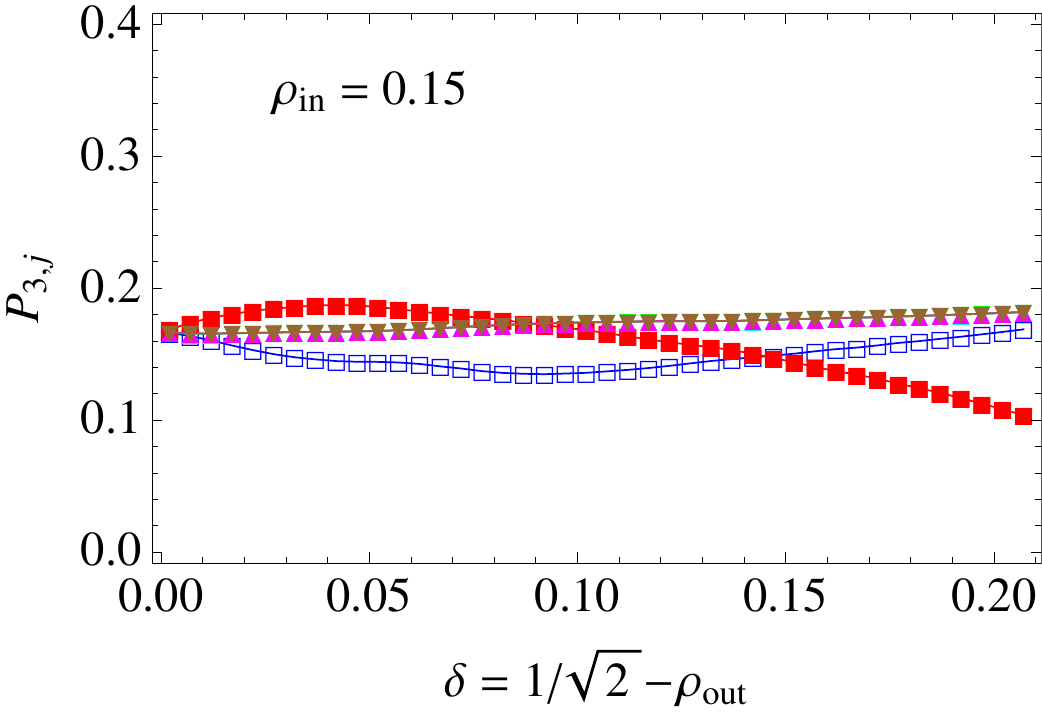}
  \label{fig.p23_b15}
  }
  \hfill
  \subfigure[]{
    \includegraphics[width=.3\textwidth,angle=0]{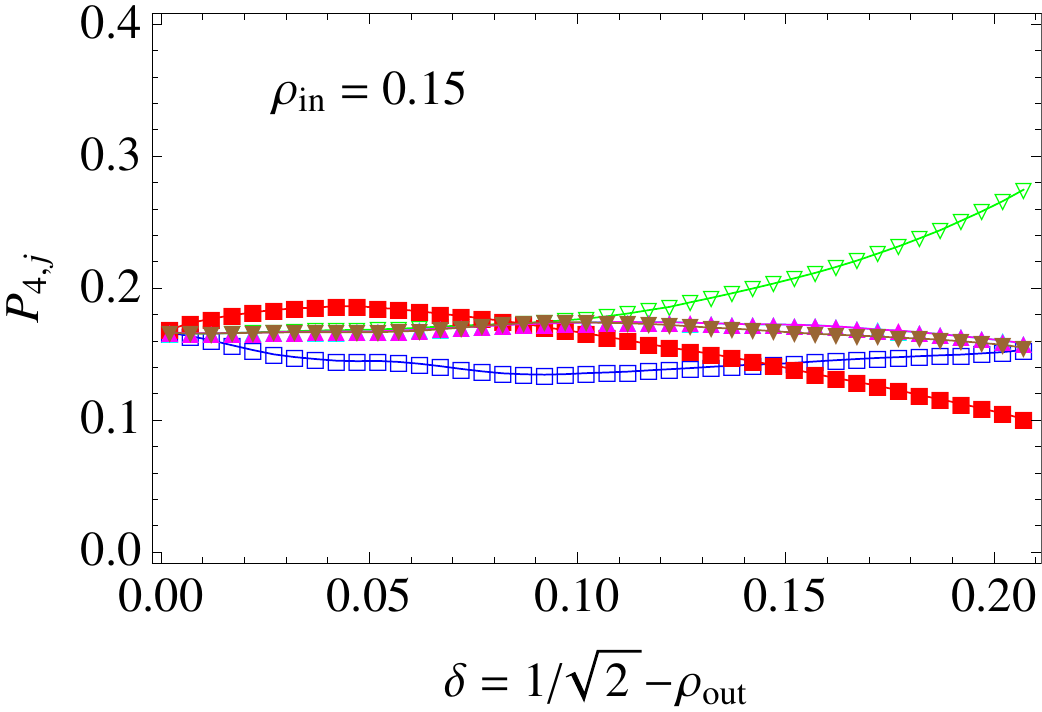}
  \label{fig.p24_b15}
  }
  \hfill
  \subfigure[]{
    \includegraphics[width=.3\textwidth,angle=0]{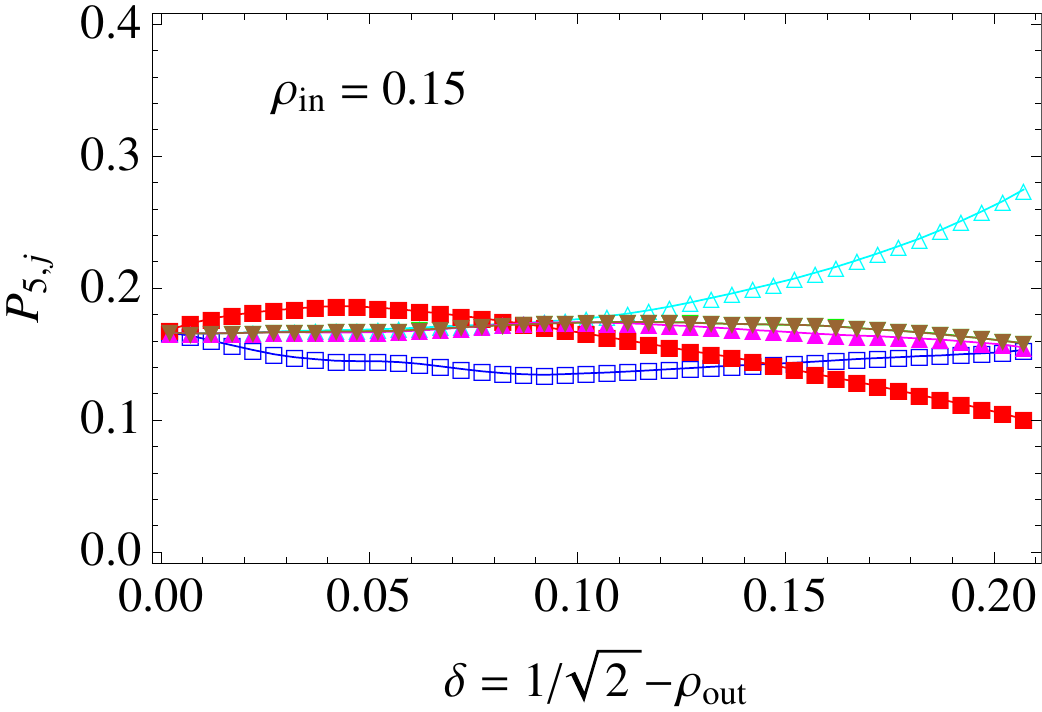}
  \label{fig.p25_b15}
  }
  \caption{Numerical computations of the probabilities 36 $P_{i,j}$
    which appear in \eref{K}, corresponding to inner radius $\rin =
    0.15$. Here again the symmetries of the process are reflected by the
    similarities between figures \ref{fig.p21_b15}, \ref{fig.p22_b15},
    \ref{fig.p24_b15} and \ref{fig.p25_b15}.
  }
  \label{fig.p2_b15}
\end{figure}

For the two-step process, the computation of the transition probabilities
$P_{i,j}$ is shown in \fref{fig.p2_b00} for $\rin = 0$, that is in the
absence of a sphere at the center of the cell. The six different panels each
correspond to a given $i=0,\ldots,5$. The same is shown in
\fref{fig.p2_b15} for $\rin = 0.15$.

\subsection{Diffusion coefficient of the billiard}

\begin{figure}[htbp]
  \centering
  \subfigure[]{
    \includegraphics[width=.45\textwidth,angle=0]{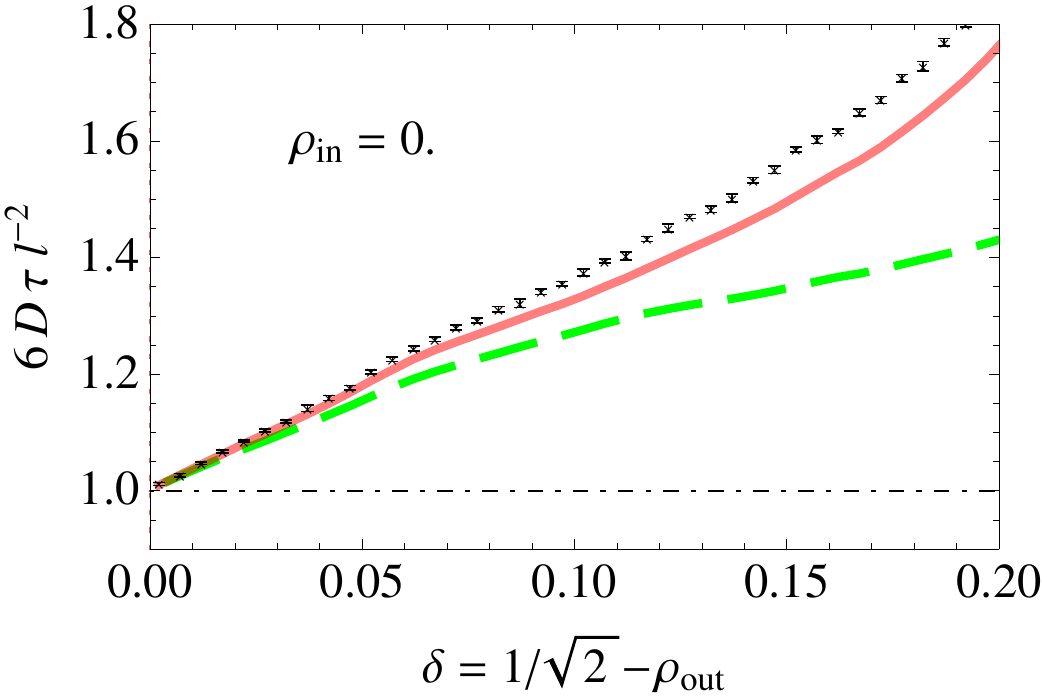}
    \label{fig.diff_b00}
  }
  \subfigure[]{
    \includegraphics[width=.45\textwidth,angle=0]{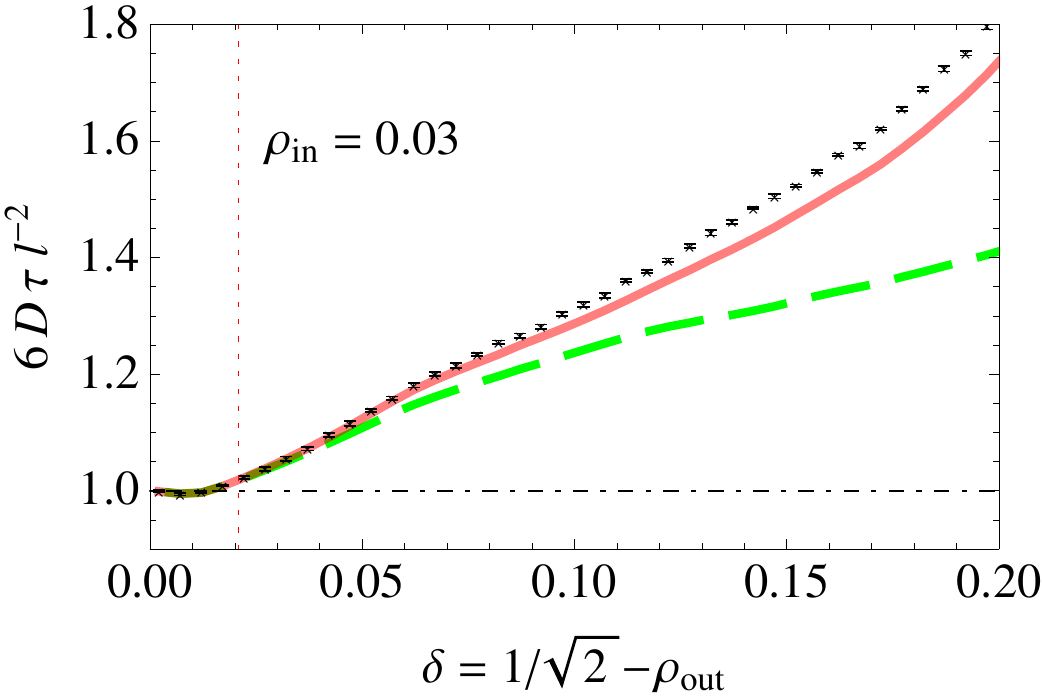}
    \label{fig.diff_b03}
  }
  \\
  \subfigure[]{
    \includegraphics[width=.45\textwidth,angle=0]{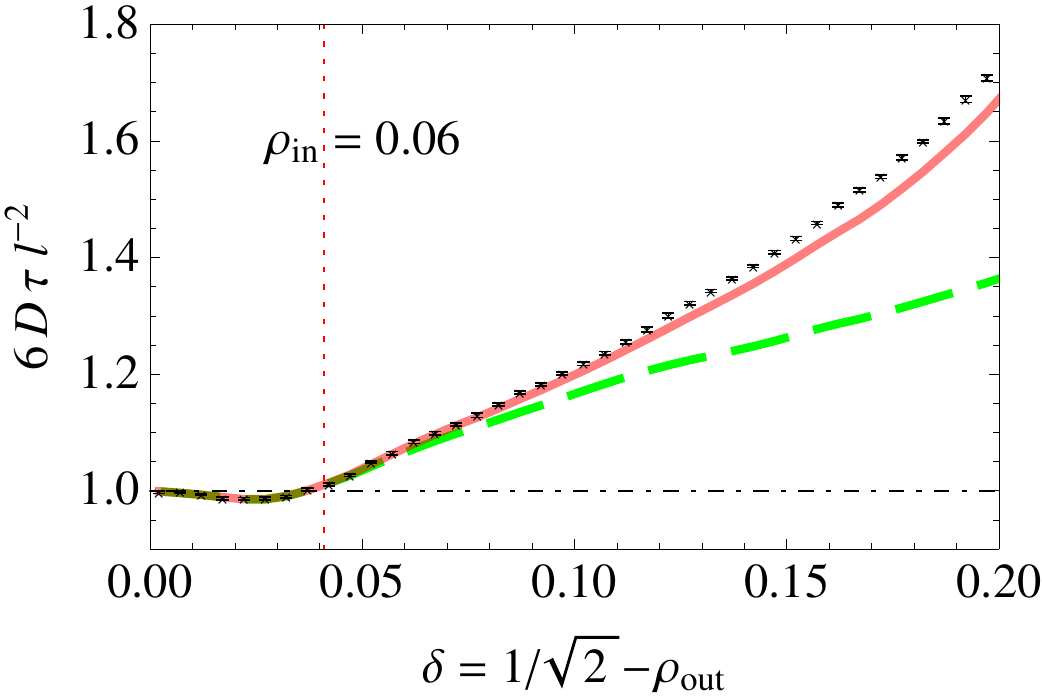}
    \label{fig.diff_b06}
  }
  \subfigure[]{
    \includegraphics[width=.45\textwidth,angle=0]{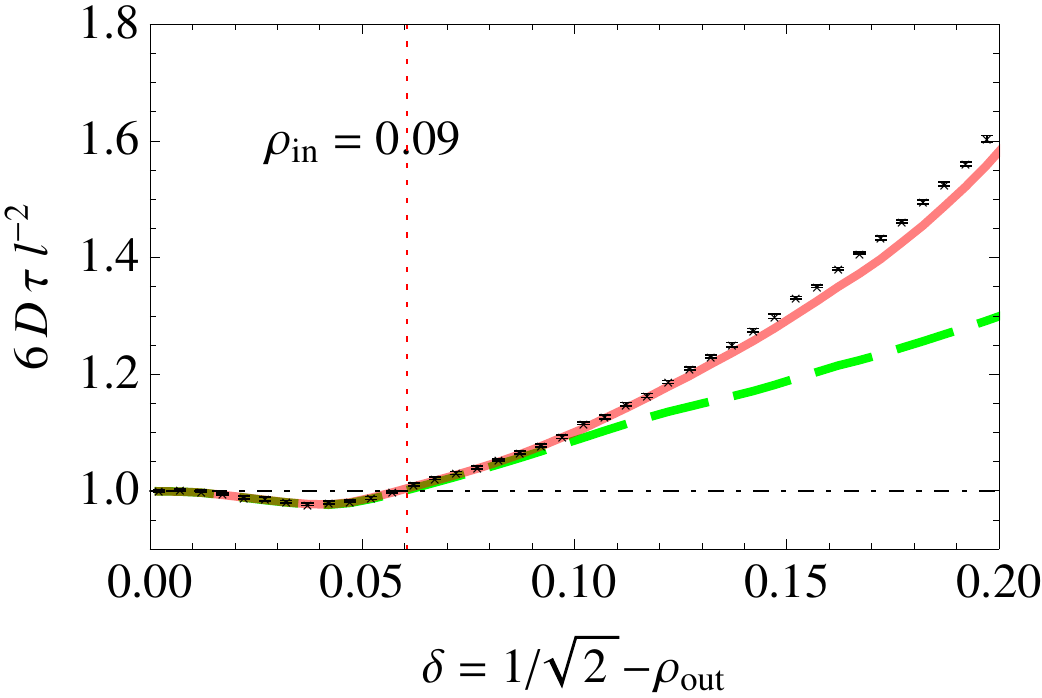}
    \label{fig.diff_b09}
  }
  \\
  \subfigure[]{
    \includegraphics[width=.45\textwidth,angle=0]{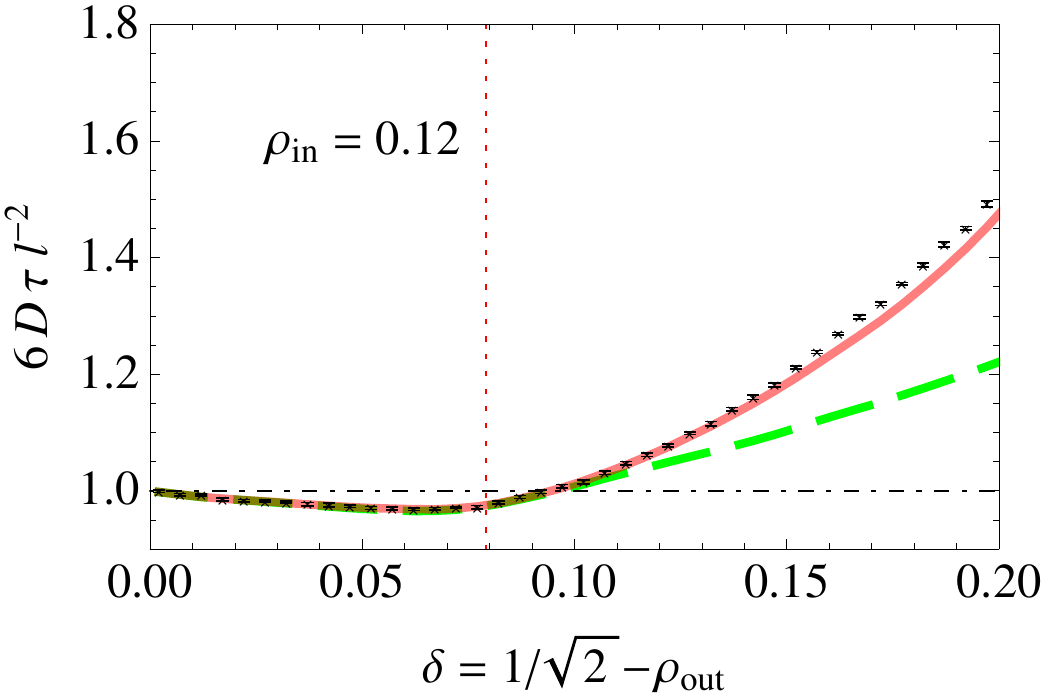}
    \label{fig.diff_b12}
  }
  \subfigure[]{
    \includegraphics[width=.45\textwidth,angle=0]{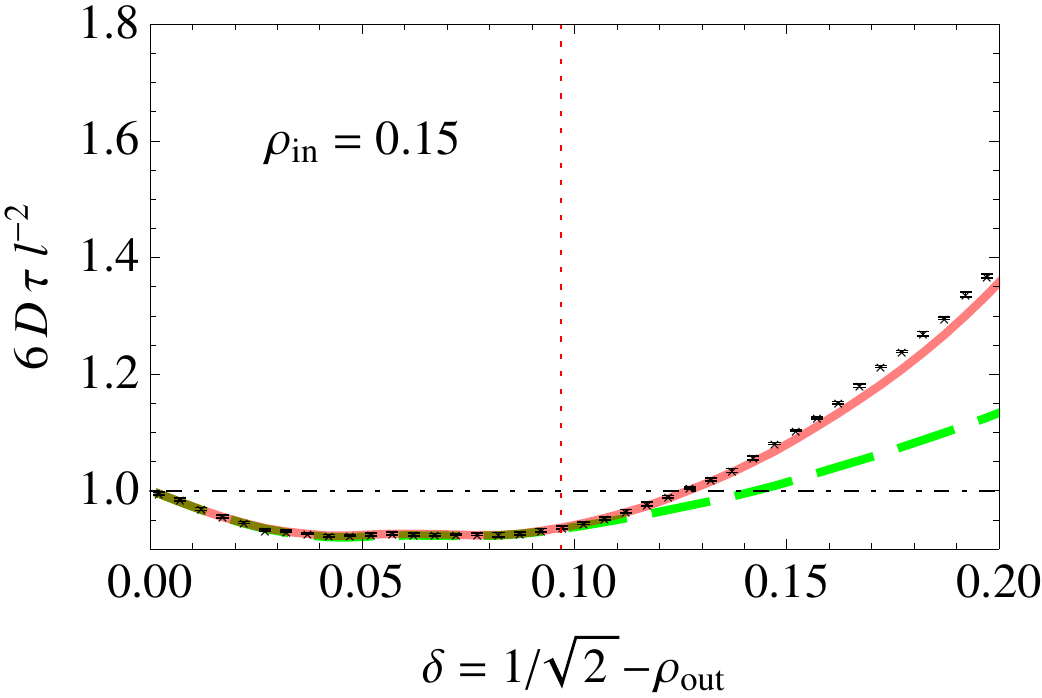}
    \label{fig.diff_b15}
  }
  \caption{Diffusion coefficient normalised with respect to the dimensional
    prediction $D_\mathrm{NMA}$, equation \eref{DNMA}, vs.~gap size  $\delta =
    1/\sqrt{2} - \rout$, plotted for different values of the inner
    radius $\rin = 0,\ldots,0.15$. The symbols (black) correspond to direct
    numerical computation of this quantity, the long dashed (green) lines
    to the single-step memory diffusion coefficient \eref{D1SMA}, and the
    solid (red) lines to the two-step memory diffusion coefficient
    \eref{D2SMA}. The vertical dotted lines indicate the separation between
    the finite and infinite horizon regimes.
  }
  \label{fig.diff}
\end{figure}

Having computed the probability transitions associated to the single and
two-step memory processes, we can compute the invariant distribution $\PP$ 
and substitute the results into equations \eref{D1SMA} and \eref{D2SMA} to
obtain values of the diffusion coefficients. These  are compared to the
diffusion coefficient of the billiard calculated from direct simulations in
\fref{fig.diff}.

We can draw several conclusions from the results shown in
\fref{fig.diff}. Firstly, we remark that in the 3D model studied here there
is relatively little back-scattering, i.e.\ motion in which the particle
reverses its direction between arriving and leaving a given cell. This
gives an important contribution to the diffusion coefficient, and, in
particular, corresponds to the fact that here we find that the diffusion
coefficient is larger than the memoryless (NMA) approximation, while in
reference \cite{Gilbert:2009p3207} the diffusion coefficient tended to lie
below the results of this approximation. Note, however, that this effect
depends strongly on the particular model used. 

It is also interesting to note that in the finite-horizon regime, i.e.\
left of the dotted vertical lines in figures
\ref{fig.diff_b03}-\ref{fig.diff_b15}, approximating  
the diffusion coefficient by the one-step memory process
\eref{D1SMA} is just as good as the two-step process
\eref{D2SMA}. In the cylindrical-horizon regime, however, the
two results are different; the single-step approximation gets
poorer as $\rout$ decreases, whereas the two-step process yields more
accurate estimates.  This corresponds to the fact that correlations decay
more slowly in the cylindrical-horizon regime 
\cite{ Sanders:2008p453}, so that memory effects persist for longer.

\section{Conclusions \label{sec.con}}

The cyclic structures of certain regular lattices underly symmetries of
their statistical properties which can be exploited to greatly simplify
their analysis. Examples are two-dimensional lattices such
as the square, the honeycomb and the triangular lattice, which were studied
in reference~\cite{Gilbert:2010p3733}. Other examples include, in higher
dimensions, the hypercubic lattices studied in this paper. Having exhibited
the cyclic structures of these lattices, we were able to extend our
previous results to hypercubic lattices with suitable adaptations, in order
to calculate the diffusion coefficients of persistent random walks with up
to two steps of memory. 

Our method is especially useful to compute the correlations of persistent
walks on such regular lattices. In particular, the velocity
autocorrelations of a two-step persistent walk may be recast in terms of
matrix powers, which can then easily be resummed to obtain a
readily-computable expression for the diffusion coefficient. 

Among the many applications of persistent random walks, deterministic
diffusive processes are ideal candidates to apply
our method. The three-dimensional periodic Lorentz gas is particularly
interesting as it exhibits two distinct types of diffusive regimes, one
with finite horizon, where memory effects decay fast, and another
with cylindrical horizon, where memory effects can remain important. In
this latter case, the approximation of the diffusive process by a two-step
memory walk proves much more accurate than the single-step process.

We remark that the application of our formalism to the diffusive properties
of Lorentz gases relies on the numerical computation of the transition 
probabilities corresponding to the persistent process with which
we approximate the deterministic process. Since there are $30$ transition
probabilities for the two-step memory walk, their analytical calculation is a daunting task.
It relies on knowledge of the statistics of trapped trajectories and involves
contributions from different time scales. Nonetheless, this computation is
formally possible, and is in principle  much simpler than that of the
actual diffusion coefficient. 

\ack

This research benefited from the joint support of FNRS
(Belgium) and CONACYT (Mexico) through a bilateral collaboration
project. The work of TG is financially supported by the Belgian Federal
Government under the Inter-university Attraction Pole project NOSY
P06/02. TG is financially supported by the Fonds de la Recherche
Scientifique F.R.S.-FNRS.  DPS acknowledges financial support from
DGAPA-UNAM grant  IN105209 and CONACYT grant CB101246.

\end{document}